# New NanoSIMS Multielement Isotope Data Reveal CO Novae As Key Sources Of $^{13}$C-rich Presolar Silicon Carbide Grains

Jordi José[1,2*†], Nan Liu[3†], Conel M. O'D. Alexander[4†], Jianhua Wang[4†]

[1*]Departament de Física, EEBE, Universitat Politècnica de Catalunya, Av. Eduard Maristany 16, Barcelona, E-08019, Spain.

[2]Institut d'Estudis Espacials de Catalunya, C. Esteve Terradas 1, Castelldefels, E-08860, Spain.

[3]Institute for Astrophysical Research, Boston University, 725 Commonwealth Av., Boston, 02215, Massachusetts, USA.

[4]Earth and Planets Laboratory, Carnegie Science, 5251 Broad Branch Rd NW, Washington, 20015, DC, USA.

*Corresponding author(s). E-mail(s): jordi.jose@upc.edu;

Contributing authors: nanliu@bu.edu; calexander@carnegiescience.edu; jwang@carnegiescience.edu;

†These authors contributed equally to this work.

## Abstract

We present new multielement NanoSIMS isotopic measurements (C, N, Si, Mg-Al, Ti, and Ni) for four putative nova SiC grains and 79 AB SiC grains from the Murchison meteorite to reassess their stellar origins. High-resolution imaging and a revised Mg/Al relative sensitivity factor for SiC yield substantially improved $^{26}$Al/$^{27}$Al ratios and the most reliable multielement characterization to date for $^{13}$C-rich presolar SiC grains. To interpret these data, we computed an expanded suite of hydrodynamic CO, ONe, and recurrent nova models spanning a range of white-dwarf masses and pre-enrichment parameters. When all isotopic systems are considered together—C, N, Mg–Al, Si, Ti, and Ni—the CO nova models provide the closest and most self-consistent match to both the putative nova grains and the subset of AB grains lacking *s*-process signatures. CO novae of low- to intermediate-mass naturally reproduce the observed $^{14}$N/$^{15}$N–$^{26}$Al/$^{27}$Al trend, the Si isotope compositions of AB grains which dominantly reflect Galactic chemical evolution (GCE), and the mild Si isotope shifts in putative nova grains relative to the GCE trend defined by AB grains. In contrast, ONe and recurrent nova models fail multiple isotopic constraints simultaneously. These results demonstrate that low- to intermediate-mass CO novae (0.6–1.0 $M_\odot$) are the most plausible stellar sources of $^{13}$C-rich SiC dust lacking *s*-process signatures (1–2% of all presolar SiC), and they establish a multielement, model-anchored framework for quantifying nova contributions to the dust reservoir in the interstellar medium.

## 1 Introduction

Classical novae are a type of stellar explosion that takes place in binary star systems, featuring a white dwarf [usually composed of carbon-oxygen (CO) or oxygen-neon (ONe)] and a low-mass main-sequence star or, occasionally, a more evolved companion such as a red giant. These explosive events are characterized by a rapid increase in optical brightness over just one to two days, reaching peak luminosities between $10^4$ and $10^5$ $L_\odot$.

The binary systems that lead to nova eruptions tend to have short orbital periods (typically less than 15 hours), allowing for mass transfer from the secondary star via Roche Lobe overflow. This transferred material carries angular momentum, resulting in the formation of an accretion disk around the white dwarf. Over time, part of this material spirals inward, accumulating on the surface of the white dwarf and building up an envelope under mildly degenerate conditions until a thermonuclear runaway sets in (see [1-5] for reviews).

Nova explosions are quite frequent in the universe, ranking as the second most common type of thermonuclear stellar explosions in our galaxy, after Type I X-ray bursts. While the expected frequency of novae in the Milky Way is about 50 per year [6], their detection from Earth is hampered by interstellar dust, which causes extinction and limits our observations to roughly 5 to 10 nova eruptions annually. Unlike Type Ia supernovae (SNeIa), where the star is completely disrupted (in most cases)[1], nova explosions do not destroy either the white dwarf or the binary system involved. Consequently, classical nova events are expected to repeat over time, typically after $10^4$ to $10^5$ years. The subclass of recurrent novae, novae that have been observed in outburst more than once, exhibit much shorter recurrence times, ranging from just 1 yr (e.g., M31N 2008-12) to about 100 yr (e.g., V2487 Oph). Such short recurrence times are only possible for very massive white dwarfs, often approaching the Chandrasekhar mass limit, along with high mass-accretion rates and very hot (luminous) white dwarfs (see, e.g., [8, 9]). Whether the recurrence times form a continuous sequence, ranging from the shorter intervals seen in recurrent novae to the longer ones associated with classical novae (i.e., 1 to $10^5$ years), remains a subject of ongoing debate. Another key difference between classical novae and SNeIa is found in the velocity of the ejected material (exceeding $10^4$ km s$^{-1}$ in SNeIa, compared to several $10^3$ km s$^{-1}$ in novae), and in the amount of mass expelled (approximately 1.4 $M_\odot$ in a SNIa and only $10^{-7}$–$10^{-4}$ $M_\odot$ in a nova outburst).

At the beginning of a nova cycle, as mass accumulates on the white dwarf surface, the pressure at the base of the accreted envelope increases. The subsequent increase in temperature (by compressional heating) triggers the onset of nuclear reactions. The vast amount of energy released during this process cannot be transported purely by radiation, which causes convection to set in once superadiabatic gradients are established within the accreted envelope. This convective motion transports short-lived $\beta^+$-unstable nuclei, such as $^{13}$N, $^{14,15}$O and $^{17}$F, synthesized deep in the envelope, toward the cooler outer layers of the star. A fraction of the energy released during the decay of these nuclei is converted into kinetic energy, fueling the eventual expansion and ejection of material during the final stages of the nova explosion. The strength of the outburst is governed by four main factors: the mass and initial luminosity (or temperature) of the white dwarf, and the metallicity and mass-transfer rate from its companion star. It is worth mentioning that such thermonuclear runaways are halted by envelope expansion, unlike Type I X-ray bursts where they are quenched by fuel depletion.

Hydrodynamic simulations of novae have shown that envelopes with solar metallicity can produce outbursts that resemble "slow novae" [10, 11]. However, to reproduce the observed features that characterize a "fast nova", enrichment in CNO elements, in the range $Z_{CNO} \sim 0.2$ to 0.5, is required [12, 13]. The source of this CNO enrichment, both supported by (theoretical) models and (observational) spectroscopic data, remains debated. Two main mechanisms have been proposed: nuclear reactions operating during the explosion or mixing processes at the boundary between the outermost white dwarf layers and the envelope. The maximum temperatures reached in a nova outburst, constrained by the inferred elemental composition of the ejecta, are limited to about $4 \times 10^8$ $K$, which is too low to drive CNO breakout and the subsequent extension of the nuclear activity to heavier nuclei. Therefore, it seems unlikely that the observed enhancements in metallicity result from thermonuclear processing. A more likely explanation involves mixing processes at the boundary between the white dwarf and the accreted envelope. Several mechanisms have been investigated in this context, including diffusion-driven mixing, shear instabilities, convective overshoot [14], and mixing by gravity wave breaking on the white dwarf surface [15, 16]. Despite these efforts, none of these approaches has proven entirely successful. Notable progress has been achieved, however, by moving beyond the limitations of spherically symmetric modeling. Indeed, multidimensional simulations of mixing at the white dwarf-envelope boundary during nova outbursts have shown that (hydrodynamic) Kelvin-Helmholtz instabilities can naturally cause self-enrichment of the accreted layers with material from the underlying white dwarf, at levels that agree with observations. In particular, three-dimensional models [17-19] have shed light on the nature of the highly fragmented, chemically enriched, and inhomogeneous structures seen in nova ejecta at high resolution. These features, which align with predictions from Kolmogorov's turbulence theory, are believed to be relics of hydrodynamic instabilities that arise during the early phases of ejection. Although such inhomogeneities in the ejecta were once thought to potentially stem from observational limitations, they may instead constitute real evidence of the turbulence generated during the runaway.

[1] See [7] for a review on Type Iax supernovae, a subclass of lower velocity ejecta and lower luminosity supernovae, in which the white dwarf is not necessarily fully disrupted during the explosion.

Spectroscopic observations of novae have provided very valuable information on the composition of the ejecta and have validated the main mechanism envisaged for the explosion. Such observations, however, often provide elemental abundances only, although a limited number of isotopic ratios are also accessible through high-resolution, near-IR spectroscopy; for instance, $^{12}C/^{13}C$ ratios were obtained through the first overtone of CO in absorption [20-22]. In comparison, presolar grains isolated from primitive meteorites serve as invaluable probes for determining isotopic abundances, thus providing nondegenerate information, i.e., abundances of nuclei instead of elements to test stellar nucleosynthesis model predictions. Presolar grains are up to micron-sized dust grains that formed in the stellar winds of evolved low-mass stars and the ejecta of explosive events like novae, before the formation of the solar system (see [23] for a recent review). After their formation, presolar grains resided in the interstellar medium for up to several billion years [24] before their incorporation into the solar system. Thus, presolar grains provide us invaluable access to probing the isotopic compositions of low-mass stellar envelopes or supernova/nova ejecta. In turn, grains that are characterized by large $^{13}C$ and $^{15}N$ excesses (C-rich grains) or $^{17}O$ excesses (O-rich grains), signatures of explosive nucleosynthesis, have been proposed to have come from novae [25-28]; the former include putative nova graphite and silicon carbide (SiC) grains, and the latter putative nova oxide and silicate grains. Similar to nova SiC grains, Type AB SiC grains are also characterized by large $^{13}C$ excesses but with a wide range of $^{14}N/^{15}N$ ratios [29]. It thus raises the question whether some, if not all, of AB grains could also have originated from novae [30, 31] rather than commonly cited alternatives such as J-type C-stars and born-again AGB stars.

Earlier attempts to explain the isotopic compositions of putative nova grains using nova nucleosynthesis models encountered several problems. These include (*i*) large, unexpected dilution factors needed for the predicted ejecta to explain the grain data, (*ii*) formation of graphite and SiC grains in O-rich diluted nova ejecta, which is forbidden by thermodynamic equilibrium calculations [32], and (*iii*) the identification of ONe novae as the most likely candidates to account for the origin of some presolar (nova) SiC grains instead of CO novae, despite the latter being more common [25, 33, 34]. More recent investigations have shown that these problems could be reconciled by fine-tuning nova modeling parameters [35] and by new CO nova models [30]. A caveat is that Iliadis et al. [35] and Bose and Starrfield [30] focused mainly on C, N, and Si isotope data that are available for most of the putative nova SiC grains, O and Si isotope data for silicates, and O isotope data alone for oxides. The limited number of elements considered likely led to misleading conclusions in some cases, as showcased by the Ca and Ti isotopic signatures (e.g., $^{44}$Ca excess) of two of their most probable nova grains, which seem inconsistent with a nova origin. Thus, multielement isotope data are the key to recognizing potential nova grains and urgently needed. In addition, given the long timespans (up to 7.6 ± 2.0 Ga; [24]) and exposure to harsh physical/chemical environments of presolar grains between their stellar births and isolation in the laboratory, presolar grains suffer from chemical alterations [36-40]. The results of these processes are isotopic compositions that can have stellar, asteroidal, and terrestrial contributions, which, reduces the power of the measured isotope data in constraining their stellar sources (e.g., [41, 42]). To address these problems, in this study we collected new NanoSIMS multielement isotope data for C, N, Si, Ti, Ni and inferred initial $^{26}$Al/$^{27}$Al ratios (whenever available) for four new putative nova and 79 AB SiC grains isolated from the Murchison meteorite; all the isotope data were collected in imaging mode at high spatial resolution so that the amounts of sampled contamination were maximally suppressed during data reduction.

This paper aims primarily to explore whether the putative nova grains and some, if not all, of the AB grains from this study could have originated from novae based on the new NanoSIMS multielement isotope data. The structure of this paper is as follows: an overview on nucleosynthesis in novae, outlining the most important isotopic signatures of these cosmic events, is given in Sect. 2. The new nova models are described in Sects. 3 and 4, and the data for putative nova and AB SiC grains are presented in Sect. 5. A thorough comparison between nova theoretical predictions and grain isotopic data is given in Sect. 6. Finally, a summary of the most relevant conclusions drawn from this study is provided in Sect. 7.

## 2 Nova Nucleosynthesis Imprints

The early stages of the thermonuclear runaway in a nova outburst are governed by the proton-proton reaction chains and the cold CNO cycles. This includes reactions such as $^{12}$C($p$, $\gamma$)$^{13}$N($\beta^+$)$^{13}$C($p$, $\gamma$)$^{14}$N. In fact, $^{12}$C($p$, $\gamma$)$^{13}$N is the single, most important reaction in triggering the explosion. As the temperature increases, the timescale for proton captures onto $^{13}$N becomes shorter than the $\beta^+$-decay time. This enables the onset of the hot CNO cycles, which involve reactions like $^{13}$N($p$, $\gamma$)$^{14}$O, along with $^{14}$N($p$, $\gamma$)$^{15}$O and $^{16}$O($p$, $\gamma$)$^{17}$F. As a result, large amounts of $^{13}$N, $^{14,15}$O, and $^{17}$F are synthesized during the outburst, ultimately leading to their decay products, $^{13}$C, $^{15}$N, and $^{17}$O, respectively, in the ejected plasma (see [4] for details). These isotopes represent the most significant contributions from novae to the chemical enrichment of the Galaxy, and their enrichments in presolar grains are thus considered most diagnostic of nova origins.

The main nuclear path during nova explosions runs close to the valley of nuclear stability and is dominated by proton captures and $\beta^+$-decays. Classical novae constitute a unique class of stellar explosions since nova nucleosynthesis allows nuclear physics inputs to rely primarily on experimental information [43]. This is because (*i*) nova nucleosynthesis is quite limited as it involves roughly one hundred nuclides with mass numbers below 40, linked through a few hundred nuclear reactions and (*ii*) the range of temperatures reached during these explosions is quite modest, lying between $10^7$ K and $4 \times 10^8$ K. Extensive research has focused on pinpointing the nuclear reactions whose uncertainties most significantly influence nova nucleosynthesis. While the vast majority of reaction rates relevant to nova nucleosynthesis have been updated in recent years (up to a desired uncertainty of 20–30%), uncertainties for some key reactions, namely $^{18}$F($p$, $\alpha$)$^{15}$O, $^{25}$Al($p$, $\gamma$)$^{26}$Si, and $^{30}$P($p$, $\gamma$)$^{31}$S, still impose significant uncertainties in the predicted yields for some species [4].

A number of isotopes synthesized during classical nova explosions may emit potentially observable $\gamma$-rays (see [44] and references therein). Among these are $^{13}$N (half-life, $\tau_{1/2} \sim 9.96$ min) and $^{18}$F ($\tau_{1/2} \sim 109.8$ min), which are responsible for an early $\gamma$-ray output, particularly at energies at or below 511 keV. In contrast, longer-lived nuclei such as $^7$Be ($\tau_{1/2} = 53.1$ day) and $^{22}$Na ($\tau_{1/2} = 2.60$ yr) decay later, once the nova envelope becomes transparent to $\gamma$-rays, powering characteristic emission lines at 478 keV and 1275 keV, respectively. While $^{26}$Al is also synthesized during nova outbursts, its decay occurs on much longer timescales ($\tau_{1/2} = 7.17 \times 10^5$ yr), and therefore only its integrated contribution across the Galaxy is observable. Nonetheless, novae are thought to account for only a modest fraction, less than 20%, of the total amount of Galactic $^{26}$Al [45, 46]. Given their relatively quick formation in stellar winds and ejecta (one to tens of years), presolar grains incorporated the initial amounts of $^{26}$Al present in their progenitor stars, which subsequently decayed to $^{26}$Mg within the grains. Putative nova graphite and SiC grains were shown to carry high initial $^{26}$Al/$^{27}$Al ratios (~0.01–0.1; [25, 33, 34]), which were inferred based on their large $^{26}$Mg excesses. In comparison, AB grains exhibit a wide range of initial $^{26}$Al/$^{27}$Al ratios, which tend to decrease with decreasing $^{15}$N enrichments [31, 47]. Neon-22 excesses in primitive meteorites were found to reside in presolar graphite grain aggregates of low-density (<2.2 g/cm$^3$) [48], which were initially thought to have been caused by the decay of $^{22}$Na produced in novae [49]. Later detailed multielement isotopic studies of single grains, however, revealed that a significant fraction of the low-density graphite grains came from core-collapse supernova (CCSNe) and that the initially incorporated $^{22}$Na was thus likely produced by CCSNe instead of novae [50, 51].

Comparisons between abundances inferred from the ejecta and those predicted by nova models typically reveal strong overall consistency, with a nucleosynthesis endpoint near Ca. However, estimates derived from spectroscopic data are often restricted to elemental (atomic) abundances, limiting the direct comparison with theoretical predictions. For putative nova grains, there is extremely limited isotope data for elements heavier than Ca. Two of the grains were shown to exhibit a $^{47}$Ti or $^{49}$Ti excess that contradicts nova nucleosynthesis [33]. In particular, one of the grains from Nittler and Hoppe [33] had large $^{28}$Si and $^{49}$Ti excesses that are in line with the isotopic signatures of Type X grains of CCSN origin. Later, Liu et al. [34] reported that two C2 grains, characterized by large $^{13}$C, $^{15}$N, and $^{29,30}$Si excesses, are enriched in neutron-rich isotopes such as $^{50}$Ti, which again argues against nova nucleosynthesis and are in better agreement with CCSN models. These grain data together suggest that, like novae, CCSN nucleosynthesis can also lead to large $^{13}$C and $^{15}$N enrichments in He-rich shells (a pattern not predicted by standard CCSN nucleosynthesis models, e.g., [52]). The possibility that CCSNe can produce "nova-like" isotopic signatures [53, 54] further complicates efforts to distinguish nova from CCSN grains. In many cases, CCSN models appear to reproduce the multielement isotopic compositions of putative nova grains more readily—but largely because prior studies have relied on ad hoc, poorly constrained mixing of material from multiple CCSN shells [34, 53, 55]. Since each shell carries distinct isotopic signatures from different burning regimes, allowing random mixing greatly expands the parameter space and can yield matches that are not physically well justified. Thus, the apparent success of CCSN fits often reflects the flexibility of the mixing prescription rather than a demonstrably better astrophysical explanation. An extended set of state-of-the-art nova models is, therefore, essential to determine whether remaining discrepancies truly argue against nova origins or simply reflect limitations of present nova simulations. This need motivated the extensive suite of hydrodynamic nova models presented in Sect. 3.

## 3. Mean Isotopic Ratios in the Nova Ejecta

A representative set of 20 hydrodynamic nova models and their mean mass-averaged isotopic compositions are summarized in Table 1. All simulations were performed with *SHIVA*, a one-dimensional Lagrangian, finite-difference, time-implicit hydrodynamic code widely used for modeling nova outbursts, Type-I X-ray bursts, and sub-Chandrasekhar supernova explosions [4, 46]. *SHIVA* solves the standard differential equations for stellar evolution and employs a general equation of state that includes contributions from a degenerate electron gas, ion plasma, and radiation, along with Coulomb corrections to the electron pressure and radiative and conductive opacities in energy transport. Nuclear energy generation is computed with a reaction network containing 120 isotopes (from $^{1}$H to $^{51}$V) connected through more than 600 nuclear interactions, with rates adopted from the *STARLIB* database ([56]; Iliadis, priv. comm.). In the models, solar-composition material [57] is accreted from the companion star, typically at a rate of $2 \times 10^{-10}$ $M_{\odot}$ yr$^{-1}$, and is assumed to mix with material from the outer layers of the underlying white dwarf at prescribed mixing fractions of 25–75%. The simulations explore a wide range of white-dwarf masses (0.6–1.35 $M_{\odot}$) and assume an initial luminosity of $10^{-2}$ $L_{\odot}$. While the range of masses considered in this work is similar to that of Bose & Starrfield [30], it is important to note that a mass threshold of approximately 1.1 $M_{\odot}$ separates CO-rich from ONe-rich white dwarfs. Accordingly, and in sharp contrast to [30], where all models were assumed to be CO-rich regardless of white dwarf mass, we adopted an ONe-rich composition for the most massive white dwarfs[2]. Additional differences relative to [30] arise from different input physics details, e.g., the values adopted for the initial white dwarf luminosity and the $\alpha$ parameter (the ratio of mixing length to pressure-scale height) used to model convective transport.

The ejected envelope consists of multiple shells of varying masses, which decrease outward. The innermost envelope layers are of particular interest for two reasons. First, they undergo the most extensive nuclear processing and therefore exhibit the strongest chemical signatures of nova nucleosynthesis. Second, their larger masses make them the dominant contributors of material that later forms dust (and presolar grains), once the expanding ejecta cools and condensation begins. Both effects are naturally accounted for in the mass-averaging procedure, which assigns weights to the individual shells according to their masses. In this sense, the mass-averaged isotopic ratios reported in this section provide an integrated view of the nucleosynthesis associated with each nova model. A detailed examination of abundance gradients in the ejecta, based on the variability of isotopic ratios across individual ejected shells, will be discussed in Sect 3.5.

### *3.1. Carbon and Nitrogen Isotopes*

All nova models reported in this work are characterized by low $^{12}$C/$^{13}$C ratios, ranging approximately between 0.4 and 2 (see Table 1), relative to the initial value of 89 taken from Lodders et al. [57], which is the solar-composition reference adopted in the models.

[2]Because the exact value of this threshold, depending on whether rotation and other physical effects are included in the evolution of their stellar progenitors, is not well determined, our model grid includes an overlap at 1.15 $M_{\odot}$, for which both CO-rich and ONe-rich white dwarfs are considered.

The dramatic reduction from the initial $^{12}C/^{13}C$ ratios (210–1100 for ONe models, 6400–19000 for CO models; see Table 2) is driven by the operation of the chain $^{12}C(p, \gamma)^{13}N(\beta^{+})^{13}C$. This chain of reactions is the single most important one at the onset of the thermonuclear runaway. In fact, the product $X(^{1}H) \cdot X(^{12}C)$, in which $X(^{1}H)$ and $X(^{12}C)$ represent the mass fractions of $^{1}H$ and $^{12}C$, respectively, at the beginning of the accretion stage determines the strength of the subsequent nova explosion [58]: a decrease in this quantity delays ignition, since fewer nuclear reactions occur and, consequently, less energy is released. This delay increases the characteristic accretion timescale, which, in turn, results in a larger accreted mass and higher pressure at the base of the envelope, ultimately leading to a more violent outburst.

The $^{12}C/^{13}C$ ratio exhibits some characteristic trends when comparing models of different masses. In low-mass CO novae, the moderate peak temperatures limit the conversion of $^{12}C$ into $^{13}C$ through nuclear interactions, resulting in comparatively high $^{12}C/^{13}C$ ratios. As the white dwarf mass, and therefore the temperature at the base of the accreted envelope, increases, more $^{13}C$ is produced, driving the $^{12}C/^{13}C$ ratio downward. However, once the temperature becomes high enough for substantial proton captures on $^{13}C$ (i.e., models with $M_{wd} > 1\ M_{\odot}$), the $^{12}C/^{13}C$ ratio begins to rise again. In ONe novae, the general trend is an increase of the $^{12}C/^{13}C$ ratio as the white dwarf mass increases. However, and in contrast to the results reported by [59], where only one or two pre-enrichment levels were explored for each white-dwarf mass, for the more extensive model grid analyzed in this work no monotonic increase is observed, except in models with 25% ONe pre-enrichment. Aside from the more extended set of models reported in this work, differences with respect to the yields discussed in [59] are also due to a number of improvements made to the SHIVA code, including updated nuclear reaction rates and updated initial solar abundances (i.e., Lodders et al. [57] vs. Anders & Grevesse [60]).

**Table 1.** Mean mass-averaged isotopic ratios for 20 hydrodynamic models of classical novae[3].

| Model | $M_{WD}$ ($M_{\odot}$) | Composition | $^{12}C/^{13}C$ | $^{14}N/^{15}N$ | $^{26}Al/^{27}Al$ | $\delta^{29}Si_{28}$ (‰) | $\delta^{30}Si_{28}$ (‰) | $\delta^{46}Ti_{48}$ (‰) |
|---|---|---|---|---|---|---|---|---|
| ONe1 | 1.15 | 25% ONe | 1.06 | 1.52 | 0.125 | –755 | –118 | 0 |
| ONe2 | 1.15 | 50% ONe | 1.15 | 1.02 | 0.121 | –756 | –420 | 0 |
| ONe3 | 1.15 | 75% ONe | 1.16 | 0.85 | 0.129 | –724 | –393 | 0 |
| ONe4 | 1.25 | 25% ONe | 1.22 | 1.02 | 0.154 | –608 | 3030 | 0 |
| ONe5 | 1.25 | 50% ONe | 1.11 | 0.79 | 0.132 | –555 | 1890 | 0 |
| ONe6 | 1.25 | 75% ONe | 1.06 | 0.69 | 0.141 | –500 | 1640 | 0 |
| ONe7 | 1.30 | 25% ONe | 1.34 | 0.99 | 0.168 | –433 | 8150 | 0 |
| ONe8 | 1.30 | 50% ONe | 1.32 | 0.56 | 0.151 | –286 | 5810 | 0 |
| ONe9 | 1.30 | 75% ONe | 1.15 | 0.45 | 0.151 | –244 | 4670 | 1 |
| ONe10 | 1.35 | 25% ONe | 2.06 | 0.72 | 0.176 | 345 | 13700 | 10 |

[3]The delta notation is defined as $\delta^{i}A_{j} = [(^{i}A/^{j}A)_{grain}/(^{i}A/^{j}A)_{solar} - 1] \times 1000$, in which $^{i}A$ and $^{j}A$ denote two isotopes of element A. Solar ratios adopted are: $^{29}Si/^{28}Si = 0.051$, $^{30}Si/^{28}Si = 0.034$, and $^{46}Ti/^{48}Ti = 0.11$.

| | | | | | | | | |
|---|---|---|---|---|---|---|---|---|
| ONe11 | 1.35 | 50% ONe | 1.52 | 0.55 | 0.156 | 446 | 12900 | 2 |
| ONe12 | 1.35 | 75% ONe | 1.29 | 0.37 | 0.154 | 631 | 13900 | 3 |
| CO1 | 0.6 | 25% CO | 1.66 | 2530 | $1.2\times10^{-3}$ | 0 | –1 | 0 |
| CO2 | 0.6 | 50% CO | 1.97 | 1080 | $7.2\times10^{-4}$ | 2 | 0 | 0 |
| CO3 | 0.8 | 25% CO | 0.45 | 172 | $5.6\times10^{-2}$ | –2 | 1 | 0 |
| CO4 | 0.8 | 50% CO | 0.56 | 102 | $3.6\times10^{-2}$ | –3 | –3 | 0 |
| CO5 | 1.0 | 25% CO | 0.53 | 12.7 | 0.401 | –54 | 14 | 0 |
| CO6 | 1.0 | 50% CO | 0.43 | 13 | 0.383 | –30 | 15 | 0 |
| CO7 | 1.15 | 25% CO | 0.81 | 2.73 | 0.225 | –469 | 55 | 0 |
| CO8 | 1.15 | 50% CO | 0.64 | 3.95 | 0.431 | –225 | 114 | 0 |

The $^{14}N/^{15}N$ ratios in the nova ejecta span a much wider range. Outbursts occurring on ONe white dwarfs produce remarkably low ratios of ~0.4–1.5, compared with the initial value of 272 adopted from the solar-composition table of Lodders et al. [57] (see Table 2, for the initial N abundances adopted). CO nova models, on the other hand, yield significantly higher ratios, ranging from ~3 to ~2500, with the largest values achieved in models with the lowest-mass white dwarfs. The strong contrast in the final N isotopic ratios between CO and ONe novae arises from large differences in the dominant nuclear reaction pathways operating during the explosion and therefore constitute a powerful diagnostic for distinguishing between low-mass and massive white dwarfs in a nova explosion.

**Table 2.** Initial mass fractions of stable C and N isotopes as a function of the adopted composition (CO vs ONe white dwarfs and different degrees of pre-enrichment).

| | **ONe Models** | | | **CO Models** | |
|---|---|---|---|---|---|
| | **25%** | **50%** | **75%** | **25%** | **75%** |
| **X($^{12}$C)** | $4.03\times10^{-3}$ | $5.74\times10^{-3}$ | $7.45\times10^{-3}$ | $1.25\times10^{-1}$ | $2.49\times10^{-1}$ |
| **X($^{13}$C)** | $2.12\times10^{-5}$ | $1.42\times10^{-5}$ | $7.08\times10^{-6}$ | $2.12\times10^{-5}$ | $1.42\times10^{-5}$ |
| **X($^{14}$N)** | $6.07\times10^{-4}$ | $4.04\times10^{-4}$ | $2.02\times10^{-4}$ | $6.07\times10^{-4}$ | $4.04\times10^{-4}$ |
| **X($^{15}$N)** | $2.39\times10^{-6}$ | $1.59\times10^{-6}$ | $7.95\times10^{-7}$ | $2.39\times10^{-6}$ | $1.59\times10^{-6}$ |

The synthesis of $^{15}N$ is highly sensitive to the abundance of $^{14}N$ in the envelope (originating from both the accreted material and the fraction produced in situ through the operation of the CNO cycle). Since CO and ONe models with identical pre-enrichment contain the same initial $^{14}N$ abundance, variations in the ejecta directly reflect differences in the thermal histories of the explosions, particularly the peak temperatures ($T_{peak}$) achieved. Indeed, the higher $T_{peak}$ values characteristic of ONe novae favor more frequent proton captures on $^{14}N$ nuclei via the sequence $^{14}N(p, \gamma)^{15}O(\beta^+)^{15}N$, leading to enhanced $^{15}N$ production and consequently lower $^{14}N/^{15}N$ ratios. Furthermore, the $^{14}N/^{15}N$ ratio in the ejecta of all CO and ONe models reported in this work decreases systematically with increasing white dwarf mass.

### *3.2. Aluminum Isotopes*

Presolar SiC grains are intrinsically Mg-free [61, 62]. The Mg detected in these grains is almost exclusively monoisotopic ($^{26}Mg$), indicating a radiogenic origin from $^{26}Al$ decay. Therefore, grain data can be directly compared with nova model predictions by examining the $^{26}Al/^{27}Al$ ratios present in the ejecta. In contrast, in some other presolar grain phases like spinel ($MgAl_2O_4$), the large presence of intrinsic Mg makes it impossible to distinguish between radiogenic and nonradiogenic $^{26}Mg$ excess [26, 63].

The synthesis of $^{26}Al$ in novae is strongly affected by the existence of a short-lived spin-isomer, $^{26}Al^m$ ($\tau_{1/2} \sim 6.35$ s). In novae, production of the longer-lived ground-state, $^{26}Al^g$ ($\tau_{1/2} \sim 7.17 \times 10^5$ yr) proceeds through proton-capture reactions on $^{25}Mg$, which populate both the ground and isomeric states; therefore, the initial abundance of $^{25}Mg$ is critical in determining the final $^{26}Al$ yield. Other nuclei such as $^{24}Mg$, $^{23}Na$, and to a lesser extent $^{20}Ne$, contribute as well to $^{26}Al$ synthesis [64]. Production pathways for $^{27}Al$ are similarly complex. Although $^{27}Al$ is primarily destroyed by a single reaction, $^{27}Al(p, \gamma)$, multiple competing channels contribute to its synthesis. A major source of $^{27}$Al is $^{26}$Mg($p$, $\gamma$), with $^{26}$Mg originating both from the amount initially present in the envelope and from the decay of $^{26}$Al$^{m}$ via $^{25}$Mg($p$, $\gamma$) or through sequential proton captures on $^{24}$Mg that lead to the formation of the $\beta^{+}$-unstable $^{26}$Si, which decays to $^{26}$Al$^{m}$. An alternative pathway involves $^{27}$Si($\beta^{+}$)$^{27}$Al, where $^{27}$Si is predominantly synthesized through $^{26}$Al$^{g,m}$($p$, $\gamma$).

ONe nova models predict a very narrow range of $^{26}$Al/$^{27}$Al ratios, approximately 0.1–0.2. In contrast, CO models cluster around two different regimes: low-mass white dwarf models produce $^{26}$Al/$^{27}$Al ~ 0.0007–0.06, whereas models with $M_{\rm wd} > 1\ M_\odot$ achieve significantly larger values, 0.2–0.4, without overlapping any of the ONe models. Therefore, the $^{26}$Al/$^{27}$Al ratio provides a diagnostic tool for distinguishing between CO and ONe novae. The $^{26}$Al/$^{27}$Al ratio increases with increasing white dwarf mass for both CO and ONe models, and for all the adopted degrees of mixing (with one exception: the 1.15 $M_\odot$ CO model with 25% pre-enrichment).

### *3.3 Silicon Isotopes*

**Table 3.** Initial mass fractions of the stable Si isotopes as a function of the adopted composition (CO vs ONe, and different degrees of pre-enrichment).

| | **ONe Models** | | | **CO Models** | |
|---|---|---|---|---|---|
| | **25%** | **50%** | **75%** | **25%** | **75%** |
| **X($^{28}$Si)** | $5.27\times10^{-4}$ | $3.51\times10^{-4}$ | $1.76\times10^{-4}$ | $5.27\times10^{-4}$ | $3.51\times10^{-4}$ |
| **X($^{29}$Si)** | $2.77\times10^{-5}$ | $1.85\times10^{-5}$ | $9.24\times10^{-6}$ | $2.77\times10^{-5}$ | $1.85\times10^{-5}$ |
| **X($^{30}$Si)** | $1.89\times10^{-5}$ | $1.26\times10^{-5}$ | $6.31\times10^{-6}$ | $1.89\times10^{-5}$ | $1.26\times10^{-5}$ |

Silicon isotope deviations from the solar composition offer also a potential diagnostic for distinguishing between CO and ONe novae. Silicon-28 is initially the most abundant Si isotope in the envelope (see Table 3). Once the thermonuclear runaway begins, the evolution of the Si isotopes in the ONe models is governed by a series of nuclear reactions involving several Al isotopes, such as $^{26}$Al$^{g}$($p$, $\gamma$)$^{27}$Si($\beta^{+}$)$^{27}$Al($p$, $\gamma$)$^{28}$Si and $^{25}$Al($p$, $\gamma$)$^{26}$Si($\beta^{+}$)$^{26}$Al$^{m}$($p$, $\gamma$)$^{27}$Si($p$, $\gamma$)$^{28}$P($\beta^{+}$)$^{28}$Si. As the temperature rises, proton captures on $^{26}$Si and $^{27}$Si become faster than the corresponding $\beta^{+}$-decays, extending the nuclear activity to heavier species through reactions such as $^{26}$Al$^{g}$($p$, $\gamma$)$^{27}$Si($p$, $\gamma$)$^{28}$P and $^{26}$Si($p$, $\gamma$)$^{27}$P($\beta^{+}$)$^{27}$Si. A key outcome at this stage is the substantial enhancement of $^{26,27}$Si that, aside from $^{28}$Si, become the most abundant Si isotopes in the envelope. Silicon-28 significantly increases in the innermost envelope layers, reaching mass fractions above $10^{-2}$. In contrast, $^{29}$Si decreases slightly due to proton-capture reactions, whereas $^{30}$Si remains largely unaffected.

For sufficiently massive white dwarf masses ($M_{\rm wd} \sim 1.35\ M_\odot$), the thermonuclear runaway heats the envelope to peak temperatures of $T_{\rm peak} > 3\times10^{8}\ K$. Under these conditions, nuclear activity in the Si-Ca mass region is dominated by multiple ($p$, $\gamma$) and ($p$, $\alpha$) reactions, together with several $\beta^{+}$-decays. At this stage, protons acquire enough energy to penetrate the Coulomb barrier of many nuclei via quantum tunneling, most notably $^{28}$Si, which, for the first time, begins to decrease slightly at the hottest envelope shells. A reduction of $^{26,27}$Si and a noticeable enhancement of $^{29,30}$Si is also found (see [65] for details). By the end of the nova outburst, $^{30}$Si is the most overproduced Si isotope in the ejecta relative to solar. The corresponding overproduction factors (i.e., ratios between mean mass fractions in the ejecta and solar values) range from ~40–60 in the 1.15 $M_\odot$ ONe models to ~250–1000 in the 1.35 $M_\odot$ ONe models.

All CO models, regardless of the adopted white dwarf mass or mixing prescription, are characterized by close-to- or lower-than-solar $\delta^{29}Si_{28}$ values, ranging from approximately 2‰ to –500‰, and close-to-solar $\delta^{30}Si_{28}$ values between –3‰ and 100‰, reflecting a very limited nuclear activity above the CNOF-mass region. In contrast, ONe models show a much larger dispersion: lower-than-solar $\delta^{29}Si_{28}$ values between –800‰ and –200‰ are predicted for white dwarf masses up to 1.3 $M_\odot$, while moderate excesses up to 600‰ are found in the ejecta of the 1.35 $M_\odot$ models. The most striking differences, however, concern the $\delta^{30}Si_{28}$ values: models with $M_{wd}$ = 1.15 $M_\odot$ exhibit lower-than-solar $\delta^{30}Si_{28}$ values down to –400‰, whereas substantial $\delta^{30}Si_{28}$ excesses are found when more massive white dwarfs are considered. Particularly striking are the models computed with $M_{wd}$ = 1.35 $M_\odot$, which show extreme excesses of approximately 13000–14000‰. This distinctive feature provides a means of linking the origin of presolar grains with large $\delta^{30}Si_{28}$ excesses to ONe novae with massive white dwarfs. As summarized in Table 1, deviations from solar in both $\delta^{29}Si_{28}$ and $\delta^{30}Si_{28}$ increase with increasing the white dwarf mass in both CO and ONe models.

### *3.4 Titanium Isotopes*

Calcium is traditionally regarded as the endpoint of nova nucleosynthesis, implying a very limited nuclear activity beyond this mass region. Consequently, the nova ejecta is expected to exhibit only small deviations from solar in all Ti isotopic ratios, largely reflecting the initial Ti isotopic abundances adopted in the models (Table 4).

Synthesis of the stable isotopes $^{46-50}$Ti during explosive H-burning proceeds through proton-capture reactions on seed Ca and Sc nuclei. The largest deviations from solar, with $\delta^{46}Ti_{48}$ values of approximately 2–10‰, are reported for models involving very massive (1.35 $M_\odot$) white dwarfs, the only ones reaching peak temperatures above $3\times10^8$ K. In these conditions, $^{46}$Ti is mainly produced via $^{44}$Ca($p$, $\gamma$)$^{45}$Sc($p$, $\gamma$)$^{46}$Ti, while its primary destruction pathway is $^{46}$Ti($p$, $\alpha$).

**Table 4.** Initial mass fractions of several stable Ca, Sc and Ti isotopes as a function of the adopted composition (CO vs ONe, and different degrees of pre-enrichment).

| | **ONe Models** | | | **CO Models** | |
|---|---|---|---|---|---|
| | **25%** | **50%** | **75%** | **25%** | **75%** |
| **X($^{44}$Ca)** | $1.13\times10^{-6}$ | $7.55\times10^{-7}$ | $3.77\times10^{-7}$ | $1.13\times10^{-6}$ | $7.55\times10^{-7}$ |
| **X($^{45}$Sc)** | $3.16\times10^{-8}$ | $2.11\times10^{-8}$ | $1.05\times10^{-8}$ | $3.16\times10^{-8}$ | $2.11\times10^{-8}$ |
| **X($^{46}$Ti)** | $1.92\times10^{-7}$ | $1.28\times10^{-7}$ | $6.39\times10^{-8}$ | $1.92\times10^{-7}$ | $1.28\times10^{-7}$ |
| **X($^{47}$Ti)** | $1.77\times10^{-7}$ | $1.18\times10^{-7}$ | $5.89\times10^{-8}$ | $1.77\times10^{-7}$ | $1.18\times10^{-7}$ |
| **X($^{48}$Ti)** | $1.78\times10^{-6}$ | $1.19\times10^{-6}$ | $5.94\times10^{-7}$ | $1.78\times10^{-6}$ | $1.19\times10^{-6}$ |

### *3.5. Isotopic Ratios in Individual Ejected Shells*

In previous sections, we reported the mean mass-averaged isotopic ratios obtained for a set of 20 hydrodynamic nova models (Table 1). This approach provides a snapshot of the main nucleosynthetic trends associated with each nova model. However, once the ejected plasma cools to temperatures below ~1500 K, formation of solids occurs locally (within specific regions of the ejecta) rather than globally (within a mass-averaged plasma). This fact motivates an alternative approach, for completeness: an analysis based on the isotopic compositions of the individual ejected shells. Although, in principle, such an analysis may appear more realistic, it is worth emphasizing that results based solely on individual shells can also be misleading. First, this approach may yield a distorted view of the overall nucleosynthetic history, since at first glance all isotopic ratios across the envelope (see Table 5) might appear equally probable. Notably, the largest deviations from the mean are often associated with the outermost, low-mass shells, which are less likely to contribute to grain formation[4]. In most cases, the variations observed in these surface zones arise from the dynamics of the retreating convective boundary rather than from nuclear processing.

[4] In the models adopted, the mass of the envelope shells decreases outward by construction. Therefore, the outermost shells are less massive than the innermost ones.

**Table 5.** Ranges of isotopic ratios for the 20 hydrodynamic nova models, based on the compositions of individual shells ejected.

| Model | $^{12}C/^{13}C$ | $^{14}N/^{15}N$ | $^{26}Al/^{27}Al$ | $\delta^{29}Si_{28}$ (‰) | $\delta^{30}Si_{28}$ (‰) | $\delta^{46}Ti_{48}$ (‰) |
|---|---|---|---|---|---|---|
| ONe1 | [0.354,2.36] | [0.735,2.51] | [0.109,0.143] | [–760, –751] | [–414, –69] | [0,0] |
| ONe2 | [0.452,2.57] | [0.545,1.73] | [0.105,0.152] | [–765, –751] | [–660, –380] | [0,0] |
| ONe3 | [0.45,2.32] | [0.589,1.22] | [0.109,0.133] | [–740, –695] | [–580, –364] | [0,0] |
| ONe4 | [0.395,2.66] | [0.643,1.15] | [0.139,0.161] | [–653, –488] | [1890,3210] | [0,0] |
| ONe5 | [0.462,2.92] | [0.459,3.86] | [0.106,0.143] | [–620, –410] | [852,2040] | [0,0] |
| ONe6 | [0.404,2.66] | [0.469,15.1] | [0.105,0.159] | [–589, –270] | [892,1750] | [0,0] |
| ONe7 | [0.44,2.85] | [0.604,7.43] | [0.125,0.18] | [–631,69] | [6520,8350] | [0,0] |
| ONe8 | [0.41,3.19] | [0.388,0.60] | [0.121,0.169] | [–487,245] | [3790,6130] | [0,0] |
| ONe9 | [0.346,2.94] | [0.394,0.483] | [0.114,0.175] | [–481,370] | [3250,4880] | [0,2] |
| ONe10 | [0.959,3.22] | [0.525,0.785] | [0.11,0.187] | [–631,2950] | [13000,15000] | [10,15] |
| ONe11 | [0.671,3.55] | [0.289,2.32] | [0.101,0.167] | [–720,4160] | [11700,14300] | [2,2] |
| ONe12 | [0.478,3.47] | [0.264,0.558] | [$7.91\times10^{-2}$,0.172] | [–620,4400] | [12400,14800] | [2,3] |
| CO1 | [1.63,2.54] | [740,4570] | [$5.1\times10^{-4}$,$1.3\times10^{-3}$] | [0,0] | [–1, –1] | [0,0] |
| CO2 | [1.77,5.04] | [419,2410] | [$1.6\times10^{-4}$,$8.5\times10^{-4}$] | [2,2] | [0,0] | [0,0] |
| CO3 | [0.411,0.749] | [37.9,716] | [$2.2\times10^{-2}$,$6.2\times10^{-2}$] | [–4,–2] | [–3,2] | [0,0] |
| CO4 | [0.456,2.17] | [35,623] | [$5.82\times10^{-3}$,$4.3\times10^{-2}$] | [–6,2] | [–3,0] | [0,0] |
| CO5 | [0.432,0.559] | [4.05,87.2] | [0.263,0.408] | [–61,–22] | [6,17] | [0,0] |
| CO6 | [0.375,1.49] | [6.28,65.7] | [0.148,0.408] | [–38,–9] | [2,17] | [0,0] |
| CO7 | [0.379,0.976] | [1.78,8.13] | [0.195,0.303] | [–504,–258] | [40,78] | [0,0] |
| CO8 | [0.477,1.5] | [3.14,8.52] | [0.399,0.453] | [–258,–88] | [60,121] | [0,0] |

While Table 5 aims primarily to facilitate a better comparison between classical nova models and presolar SiC grains of potential nova origin (see Sect. 6), given that grains likely condensed locally within the ejecta, several additional aspects are worth noting. The shell-by-shell analysis reveals an isotopically diverse ejecta in which layers with $^{12}C > ^{13}C$ (or $^{14}N > ^{15}N$) coexist alongside others with $^{12}C < ^{13}C$ (or $^{14}N < ^{15}N$), in some cases. The $^{26}Al/^{27}Al$ ratios display very little dispersion across the ejecta for all models, reinforcing its potential as a diagnostic for distinguishing between CO and ONe novae. Finally, the $\delta^{29,30}Si_{28}$ values show that, in some models (e.g., ONe7 – ONe12, CO3 – CO4), portions of the ejecta exhibit lower-than-solar compositions while other shells simultaneously show $^{29,30}Si$ excesses. Therefore, conclusions based exclusively on mean, mass-averaged ratios (e.g., all ONe models with massive white dwarfs, $M_{wd} = 1.35\ M_{\odot}$, are characterized by $\delta^{29}Si_{28} > 0$; see Table 5) must be regarded with caution.

## 4. Isotopic Ratios in Recurrent Novae

As noted in previous sections, Ca is traditionally considered the endpoint of nova nucleosynthesis. However, because some of the AB grains analyzed in this study (Section 5) exhibit moderate excesses in Ti, it is worthwhile to also consider predictions from recurrent novae, a subclass of novae with potential nuclear activity in the Si–Ti mass region. Tables 6 and 7 summarize the isotopic ratios in the ejecta for 14 recently computed hydrodynamic models[5] [8] developed in the framework of the forthcoming outburst of T Coronae Borealis, a recurrent nova with a ~80-yr recurrence period whose next eruption was originally predicted for 2025 [66, 67]. To date, no presolar grain study has explicitly explored the potential contribution of recurrent novae.

About a dozen recurrent novae have been discovered in the Milky Way (with additional systems identified in nearby galaxies, such as Andromeda). Their short recurrence periods between successive eruptions, typically 1 yr to 100 yr, can only be reproduced numerically when simultaneously adopting high white dwarf masses, high initial luminosities (or temperatures), and high mass-accretion rates. Recurrent novae generally eject less mass than classical novae ($10^{-7}$–$10^{-6}\ M_{\odot}$ versus $10^{-5}$–$10^{-4}\ M_{\odot}$) and usually exhibit solar metallicity ejecta, implying little to no mixing between the accreted envelope and the outermost layers of the white dwarf. Consequently, most of the models listed in Tables 6 and 7 assume accretion of material with solar composition. However, models RN13 and RN14 were computed using $Z = 0.1\ Z_{\odot}$ and $10\ Z_{\odot}$, respectively, to quantitatively assess the influence of the metallicity of the accreted material on the ejecta composition.

[5]No mixing with the underlying white dwarf substrate is assumed for recurrent nova models, consistent with observations indicating solar-composition ejecta. Consequently, the specific composition of the white dwarf, whether CO- or ONe-rich, is not relevant in this context.

**Table 6.** Mean mass-averaged isotopic ratios for a series of 14 hydrodynamic models of recurrent novae.

| Model | $M_{WD}$ ($M_\odot$) | $L_{WD}$ ($L_\odot$) | $Z_{acc}$ ($Z_\odot$) | $^{12}C/^{13}C$ | $^{14}N/^{15}N$ | $^{26}Al/^{27}Al$ | $\delta^{29}S_{28}$ (‰) | $\delta^{30}Si_{28}$ (‰) | $\delta^{46}Ti_{48}$ (‰) |
|---|---|---|---|---|---|---|---|---|---|
| RN1 | 1.20 | 1.0 | 1.0 | 0.757 | 48.5 | 0.101 | –872 | 173 | 0 |
| RN2 | 1.20 | 0.1 | 1.0 | 0.692 | 22.1 | 0.155 | –890 | 682 | 0 |
| RN3 | 1.20 | 0.01 | 1.0 | 0.774 | 4.57 | 0.183 | –870 | 6630 | 0 |
| RN4 | 1.30 | 1.0 | 1.0 | 0.644 | 6.83 | 0.185 | –882 | 4120 | 0 |
| RN5 | 1.30 | 0.1 | 1.0 | 0.678 | 4.82 | 0.191 | –876 | 7130 | 0 |
| RN6 | 1.30 | 0.01 | 1.0 | 0.807 | 2.58 | 0.194 | –862 | 18200 | 0 |
| RN7 | 1.35 | 1.0 | 1.0 | 0.777 | 1.58 | 0.200 | –850 | 26200 | 0 |
| RN8 | 1.35 | 0.1 | 1.0 | 0.935 | 1.45 | 0.197 | –837 | 23600 | 9 |
| RN9 | 1.35 | 0.01 | 1.0 | 1.30 | 1.31 | 0.188 | –756 | 21600 | 436 |
| RN10 | 1.38 | 1.0 | 1.0 | 0.900 | 1.23 | 0.198 | –816 | 35300 | 103 |
| RN11 | 1.38 | 0.1 | 1.0 | 1.14 | 1.16 | 0.187 | –764 | 25600 | 473 |
| RN12 | 1.38 | 0.01 | 1.0 | 1.44 | 1.11 | 0.193 | –599 | 21000 | 1780 |
| RN13 | 1.38 | 0.1 | 0.1 | 1.29 | 1.14 | 0.174 | –93 | 10900 | 130000 |
| RN14 | 1.38 | 0.1 | 10.0 | 0.552 | 1.15 | 0.196 | –828 | 5880 | 0 |

**Table 7.** Ranges of isotopic ratio variations for the 14 hydrodynamic recurrent nova models based on individual shells ejected.

| Model | $^{12}C/^{13}C$ | $^{14}N/^{15}N$ | $^{26}Al/^{27}Al$ | $\delta^{29}Si_{28}$ (‰) | $\delta^{30}Si_{28}$ (‰) | $\delta^{46}Ti_{48}$ (‰) |
|---|---|---|---|---|---|---|
| RN1 | [0.599,0.792] | [17.9,107] | [0.098,0.105] | [–872, –871] | [171,174] | [0,0] |
| RN2 | [0.481,0.765] | [9.28,36.9] | [0.147,0.167] | [–891, –889] | [677,684] | [0,0] |
| RN3 | [0.355,1.22] | [1.78,7.24] | [0.176,0.188] | [–882, –859] | [6540,6660] | [0,0] |
| RN4 | [0.314,0.812] | [2.5,14.5] | [0.182,0.190] | [–883, –879] | [4100,4130] | [0,0] |
| RN5 | [0.302,0.9] | [1.77,9.63] | [0.188,0.194] | [–878, –872] | [7090,7150] | [0,0] |
| RN6 | [0.35,1.17] | [1.06,5.51] | [0.186,0.206] | [–877, –849] | [18100,18200] | [0,0] |
| RN7 | [0.266,1.18] | [0.724,4.66] | [0.19,0.219] | [–860, –842] | [26200,26300] | [0,0] |
| RN8 | [0.312,1.39] | [0.763,3.94] | [0.181,0.21] | [–843, –826] | [23600,23800] | [9,9] |
| RN9 | [0.651,1.53] | [0.742,3.02] | [0.148,0.201] | [–769, –734] | [20500,22300] | [436,440] |
| RN10 | [0.234,1.39] | [0.73,3.39] | [0.171,0.210] | [–820, –808] | [35100,35600] | [103,103] |
| RN11 | [0.419,1.50] | [0.748,3.04] | [0.14,0.205] | [–774, –746] | [25000,26000] | [471,475] |
| RN12 | [0.73,2.1] | [0.741,2.57] | [0.12,0.215] | [–652, –422] | [16900,22000] | [1770,1790] |
| RN13 | [1.14,1.87] | [0.666,8.33] | [0.115,0.194] | [–731,2370] | [10100,13100] | [129000,130000] |
| RN14 | [0.187,0.985] | [0.187,0.985] | [0.183,0.21] | [–853, –814] | [5790,5960] | [0,0] |

### *4.1 Recurrent Nova Models of Solar Metallicity*

In this section, we analyze the results obtained for the set of hydrodynamic models RN1 through RN12, corresponding to accretion of material with solar metallicity ($Z_\odot$). The resulting $^{12}C/^{13}C$ ratios range from 0.6 to 1.4, partially overlapping with those reported for (classical) ONe novae. For recurrent novae, the ratio generally increases with the white dwarf mass, with the exception of the two 1.3 $M_\odot$ models computed with $L_{wd}$ = 0.1 and 1 $L_\odot$. In all cases, $^{12}C/^{13}C$ reaches its highest value for $L_{wd}$ = 0.01 $L_\odot$, regardless of the adopted $M_{wd}$.

The $^{14}N/^{15}N$ ratios range from 1.1 to 49, with minimal overlap with those reported for ONe novae. The ratio decreases monotonically with increasing $M_{wd}$, regardless of the adopted $L_{wd}$. In addition, $^{14}N/^{15}N$ decreases with decreasing $L_{wd}$, for any given $M_{wd}$.

The $^{26}Al/^{27}Al$ ratios exhibit little dispersion, spanning 0.1 to 0.2, fully overlapping with the values characteristic of ONe novae. In models computed with $L_{wd}$ = 0.1 and 1 $L_\odot$, $^{26}Al/^{27}Al$ tends to increase with $M_{wd}$, reaching a maximum for 1.35 $M_\odot$. For models computed with $L_{wd}$ = 0.01 $L_\odot$, the maximum is instead found for 1.3 $M_\odot$. No clear dependence of $^{26}Al/^{27}Al$ on $L_{wd}$ is found for a given $M_{wd}$, except that in the 1.2 and 1.3 $M_\odot$ models the ratio increases as $L_{wd}$ decreases.

All recurrent nova models reported in this study yield lower-than-solar $\delta^{29}Si_{28}$ values, spanning approximately –900‰ to –600‰. The $\delta^{29}Si_{28}$ value generally increases with increasing $M_{wd}$, monotonically for models computed with $L_{wd} = 0.1\ L_{\odot}$ and $0.01\ L_{\odot}$, whereas the 1 $L_{\odot}$ models reach a maximum for $M_{wd} = 1.3\ M_{\odot}$. It is also worth noting that $\delta^{29}Si_{28}$ tends to increase with decreasing $L_{wd}$, for most of the models analyzed.

With regard to $\delta^{30}Si_{28}$, all 12 models predict excesses, relative to solar, ranging from 170‰ to 35000‰. The $\delta^{30}Si_{28}$ value tends to increase with $M_{wd}$: monotonically for the 0.1 $L_{\odot}$ and 1 $L_{\odot}$ model sets, while for the 0.01 $L_{\odot}$ models the maximum occurs at 1.35 $M_{\odot}$. The dependence of $\delta^{30}Si_{28}$ on $L_{wd}$ is mass-dependent: for the 1.2 and 1.3 $M_{\odot}$ models $\delta^{30}Si_{28}$ increases with decreasing $L_{wd}$, whereas the opposite trend is found for the 1.35 $M_{\odot}$ and 1.38 $M_{\odot}$ models.

Finally, although most models predict close-to-solar Ti isotopic ratios, models with the most massive white dwarfs (i.e., 1.35 $M_{\odot}$ and 1.38 $M_{\odot}$) exhibit striking $\delta^{46}Ti_{48}$ values in the range 100–1800‰. These values have to be taken with caution, because Ti lies at the upper boundary of the nuclear network employed in these simulations; nevertheless, if confirmed, $\delta^{46}Ti_{48}$ may represent a potentially powerful diagnostic for identifying massive white dwarfs in recurrent nova systems. Future work with an expanded reaction network will be required to fully assess this possibility.

### *4.2 Recurrent Nova Models with $Z = 0.1\ Z_{\odot}$ and $10\ Z_{\odot}$*

A direct comparison among the recurrent nova models computed with 0.1 $Z_{\odot}$ (RN13), $Z_{\odot}$ (RN11), and 10 $Z_{\odot}$ (RN14) provides insight into potential dependencies on the metallicity of the accreted material (see Table 6). Increasing the metallicity of the accreted material leads to lower $^{12}C/^{13}C$, $\delta^{29}Si$ and $\delta^{46}Ti$ values, together with an enhancement in the $^{26}Al/^{27}Al$ ratio. The $^{14}N/^{15}N$ ratio remains nearly unchanged across these models, while no clear trend is evident for $\delta^{30}Si$; however, large $\delta^{30}Si$ excesses (~5900–26200‰) are consistently predicted, independent of the initial metallicity adopted.

## 5 New NanoSIMS Data for Putative Nova and AB SiC Grains

### *5.1 Methodology*

The presolar SiC grains in this study were extracted from the Murchison (CM2) carbonaceous chondrite through the CsF dissolution technique [68]. Two sample mounts were prepared, Mount #5 and #6, and SiC grains >500 nm in size were first identified by automated backscattered electron−energy dispersive X-ray particle analyses by following the previously established procedure [69]. Isotopic analyses of presolar SiC grains were carried out with the nano-scaled secondary ion mass spectrometer (NanoSIMS) 50L ion microprobe at the Earth and Planets Laboratory of the Carnegie Institution for Science, except for the C, N, and Si isotope measurements of grains from Mount #5, which were obtained with the NanoSIMS 50 at Washington University in St. Louis. NanoSIMS is a secondary ion mass spectrometer in which a focused primary ion beam ($Cs^{+}$ or $O^{-}$) sputters atoms and molecules from the sample surface; the resulting secondary ions are analyzed by a double-focusing mass spectrometer that combines a magnetic sector with an electrostatic analyzer, providing simultaneous energy and angular focusing and thus high mass resolution and transmission. The NanoSIMS 50 instrument is equipped with five detectors for the simultaneous collection of five secondary ion species, whereas the NanoSIMS 50L includes two additional detectors, enabling concurrent measurement of up to seven ion species. The measurement protocols for C, N, Si, and Mg-Al isotopes follow those in [62], for Ti isotopes those in [40], and for Ni isotopes those in [70]. Initial grain classification into mainstream, Y, Z, X, and AB types employed the standard scheme based on C, N, and Si isotopic compositions [71]. The identified AB and putative nova grains were subsequently analyzed for Mg-Al, Ti, and Ni isotopes. All measurements were performed in imaging mode using electron multipliers, with a typical spatial resolution of ~100 nm for C, N, Si, Ti, and Mg-Al isotopes and slightly poorer resolution for Ni isotopes due to the generally lower Ni concentrations that required a larger primary ion beam.

Data reduction was performed using the *L'image* software package (version 12-15-2020; https://limagesoftware.net). For each grain, regions of interest (ROIs) were defined to exclude contributions from adjacent or overlying particles, aided where possible by high-resolution secondary electron microscopic imaging. Grain edges showing signal enhancements that are indicative of contamination were avoided. Presolar SiC grains are typically enriched in Ti and Ni but may show surface contamination from Ca, Cr, Fe, and Zn. The main isobaric interferences were $^{48}$Ca on $^{48}$Ti, $^{50}$Cr on $^{50}$Ti, $^{58}$Fe on $^{58}$Ni, and $^{64}$Zn on $^{64}$Ni. The interference elements Ca, Cr, and Fe, which are especially prevalent at grain edges, were minimized through careful ROI selection.

Contributions from $^{48}$Ca, $^{50}$Cr, and $^{58}$Fe were found to be negligible owing to the high intrinsic Ti/Ca, Ti/Cr, and Ni/Fe ratios in the grains. In contrast, the $^{64}$Zn interference was more significant due to its high natural abundance (49.2%) relative to $^{64}$Ni (0.926%). For $\delta^{64}$Ni/$^{58}$Ni, only grains with <50% $^{64}$Zn contribution at mass 64 are reported in Table 1. Interference correction employed a $^{64}$Zn/$^{66}$Zn ratio determined from NIST 610 glass, and $^{64}$Ni/$^{58}$Ni ratios were calibrated using Zn-free Murchison Fe-oxide grains present on the same sample mounts. This correction is justified by the spatial distribution of Zn which is confined mostly to grain edges and its likely origin as solar-composition surface contamination. Among corrected grains, Zn contributions ranged from 0 to 50% (mean 19%). The Murchison Fe-oxide grains were enriched in Ni ($^{58}$Ni$^{+}$/$^{56}$Fe$^{+}$~10%), and the contribution of $^{58}$Fe to $^{58}$Ni was estimated to be $\lesssim$5%. This interference was corrected by subtracting the $^{58}$Fe contribution assuming a terrestrial isotopic composition. The revised Mg/Al relative sensitivity factor (RSF) for ionization by the SIMS O$^{-}$ primary ion beam from [62] was used to derive initial $^{26}$Al/$^{27}$Al ratios. Reported 1σ uncertainties include both counting statistics and analytical reproducibility but do not include the ±6% uncertainties in the Mg/Al RSF reported in [62].

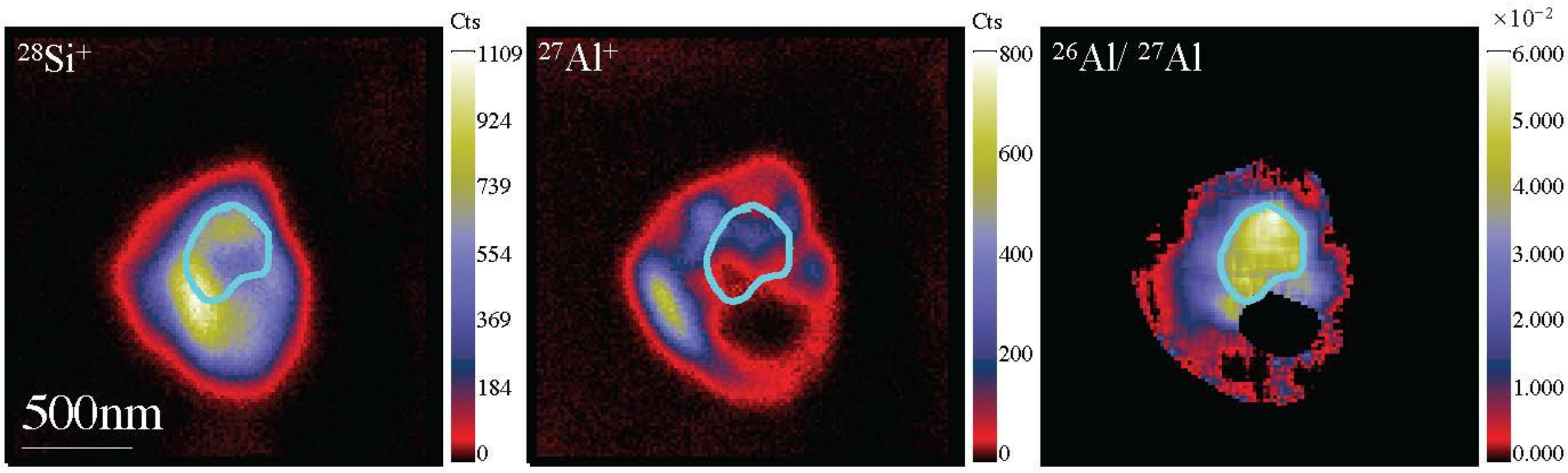

**Fig. 1**. NanoSIMS secondary ion images of nova grain M5-A6-0831. The grain had significant Al contamination concentrated at the edge, supported by the lowered inferred initial $^{26}$Al/$^{27}$Al ratios. A small ROI shown in cyan was thus used to suppress Al contamination to extract intrinsic signals.

The efficient suppression of extrinsic signals using our approach is exemplified in Fig. 1. As discussed in [40], Al contamination is common among presolar SiC grains, which was also observed at the surface of the nova grain M5-A6-0831. In addition to the surface Al contamination, the grain also contained an Al-rich subgrain, despite its general depletion of Al in the interior. By applying a small ROI centered on the subgrain, we obtained an initial $^{26}Al/^{27}Al$ ratio of $(4.99\pm0.18)\times10^{-2}$, in comparison to a ratio of $(1.85\pm0.06)\times10^{-2}$ for the entire grain. Our approach thus allowed for an efficient suppression of contamination, increasing the derived initial ratio by a factor of 2.7.

### *5.2 New NanoSIMS Isotope Data*

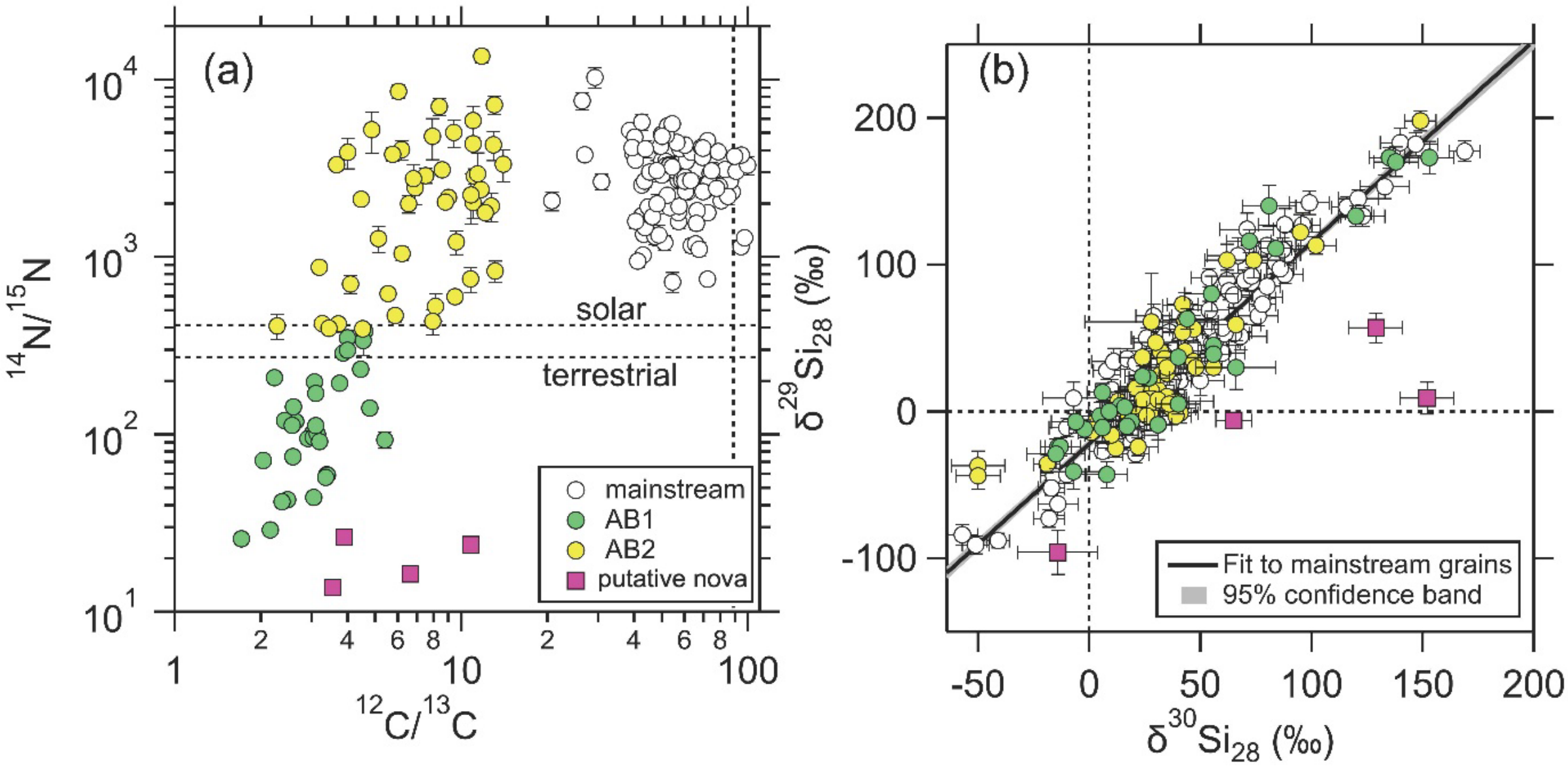

**Fig. 2**. Carbon, N, and Si isotope ratios of mainstream, AB1, AB2, and putative nova SiC grains. The dashed lines represent terrestrial compositions unless noted otherwise. The linear fit to the mainstream grain data in panel (b) was obtained by using the CEREsFit.xlsm tool of [72] and follows the equation of $\delta^{29}Si_{28} = (1.49\pm0.03) \times \delta^{30}Si_{28} - (26.5\pm1.8)$ with 1σ errors and a mean squared weighted deviation (MSWD) value of 2.2. An MSWD value of unity means a perfect linear correlation.

5.2.1 Carbon, N, and Si isotope data

All comparisons and statistical analyses below are based exclusively on the data obtained in this work under the same analytical conditions to ensure internal consistency. While previous studies (e.g., [31, 73]) have combined literature datasets, such compilations inherently include measurements acquired under different instrumental conditions and analytical protocols, which can introduce additional scatter and systematic offsets. Because the distinctions explored here between grain populations are relatively subtle, we restrict our analysis to a self-consistent dataset to minimize these effects and ensure robust comparisons. In the following discussions, we first discuss AB grains as a whole to establish their overall characteristics and then subdivide them into AB1 and AB2 where statistically meaningful differences are observed. For Ti and Ni isotopes, AB1 and AB2 grains are not discussed separately (but plotted separately for visual inspection) because no statistically significant differences are resolved between the two subgroups within analytical uncertainties.

The four new putative nova and 79 new AB grains are shown in Fig. 2, in which 102 mainstream grains that were analyzed under the same conditions [70] were also included for direct comparison. This is because mainstream grains, the dominant population of presolar SiC grains (>85%), are a group of presolar grains with the best-known stellar origin, thanks to their extensive isotopic studies over the years (see [23] for a recent review). The current consensus is that mainstream SiC grains came from low-mass (1.5–4 $M_{\odot}$) AGB stars with close-to- and higher-than-solar metallicities [74, 75]. Mainstream grains are defined to have $^{12}C/^{13}C$ ratios lying between 20 and 100 and characterized by large $^{14}N$ excesses [40]. In addition, the varying Si isotope ratios among mainstream grains in Fig. 2b are primarily caused by the varying initial stellar compositions of their parent AGB stars, reflecting Galactic chemical evolution (GCE). This is because state-of-the-art low-mass AGB stellar nucleosynthesis models predict quite small changes to the initial Si isotopic compositions by the slow neutron-capture process (*s*-process) at close-to and higher-than solar metallicities [74, 76].

In comparison to mainstream grains, AB grains exhibit larger $^{13}C$ enrichments ($1.7 \leq {}^{12}C/^{13}C \leq 14.0$) and a much wider range of $^{14}N/^{15}N$ ratios (Fig. 2a). Type AB grains, however, generally follow the same Si GCE trend as mainstream grains within 1σ uncertainties (Fig. 2b), suggesting that their Si isotopic compositions also mainly reflect initial stellar compositions and were minimally altered by nucleosynthesis in their parent stars. In addition, on average, AB grains show slightly smaller $^{29,30}Si$ excesses, relative to $^{28}Si$, than mainstream grains (Fig. 2); for instance, the median $\delta^{29}Si_{28}$ and $\delta^{30}Si_{28}$ values are 14‰ and 39‰, respectively, for AB grains, and 46‰ and 42‰, respectively, for mainstream grains, with the corresponding standard deviations being comparable between the two groups. The smaller $\delta^{29}Si_{28}$ values of AB grains imply that their parent stars had lower initial stellar metallicities than those of mainstream grains.

AB grains were proposed to be further divided into two subgroups, AB1 and AB2 grains based on their $^{14}N/^{15}N$ ratios (Fig. 2a; [31]). In line with the observation reported in [34], the new AB1 and AB2 grains analyzed here were found to follow slightly different Si GCE trends: $\delta^{29}Si_{28} = (1.41\pm0.05) \times \delta^{30}Si_{28} - (16.8\pm3.1)$ for AB1 grains and $\delta^{29}Si_{28} = (1.55\pm0.08) \times \delta^{30}Si_{28} - (29.2\pm3.4)$ for AB2 grains (difference <2σ significance). Our observed AB1-AB2 differences in the slope and intercept are in line with the trends observed in [31], but the differences observed here are reduced. This is likely because Liu et al. [31] included literature AB grain data in addition to their new data when obtaining the fit, leading to systematic uncertainties in their results introduced by the different analytical conditions. In summary, our new data support that AB1 and AB2 grains, exhibiting different ranges of $^{14}N/^{15}N$ ratio, follow slightly different Si GCE trends.

Our new data also, for the first time, reveal a significant distinction between AB1 and putative nova grains in their C and N isotope ratios, in addition to Si isotopes (Fig. 2a). The new data show that AB1 grains tend to exhibit larger $^{13}C$ excesses with increasing $^{15}N$ enrichments, pointing to the effect of hot CNO burning; in contrast, putative nova grains tend to show smaller $^{13}C$ excesses with increasing $^{15}N$ enrichments. In line with previous observations [25, 34], the new putative nova grains deviate from the mainstream Si GCE line toward larger $^{30}Si$ excesses.

5.2.2 Aluminum-26

**Fig. 3**. Plots of inferred initial $^{26}Al/^{27}Al$ ratios versus $^{14}N/^{15}N$ for grains from this study. The mainstream grain data are from [70].

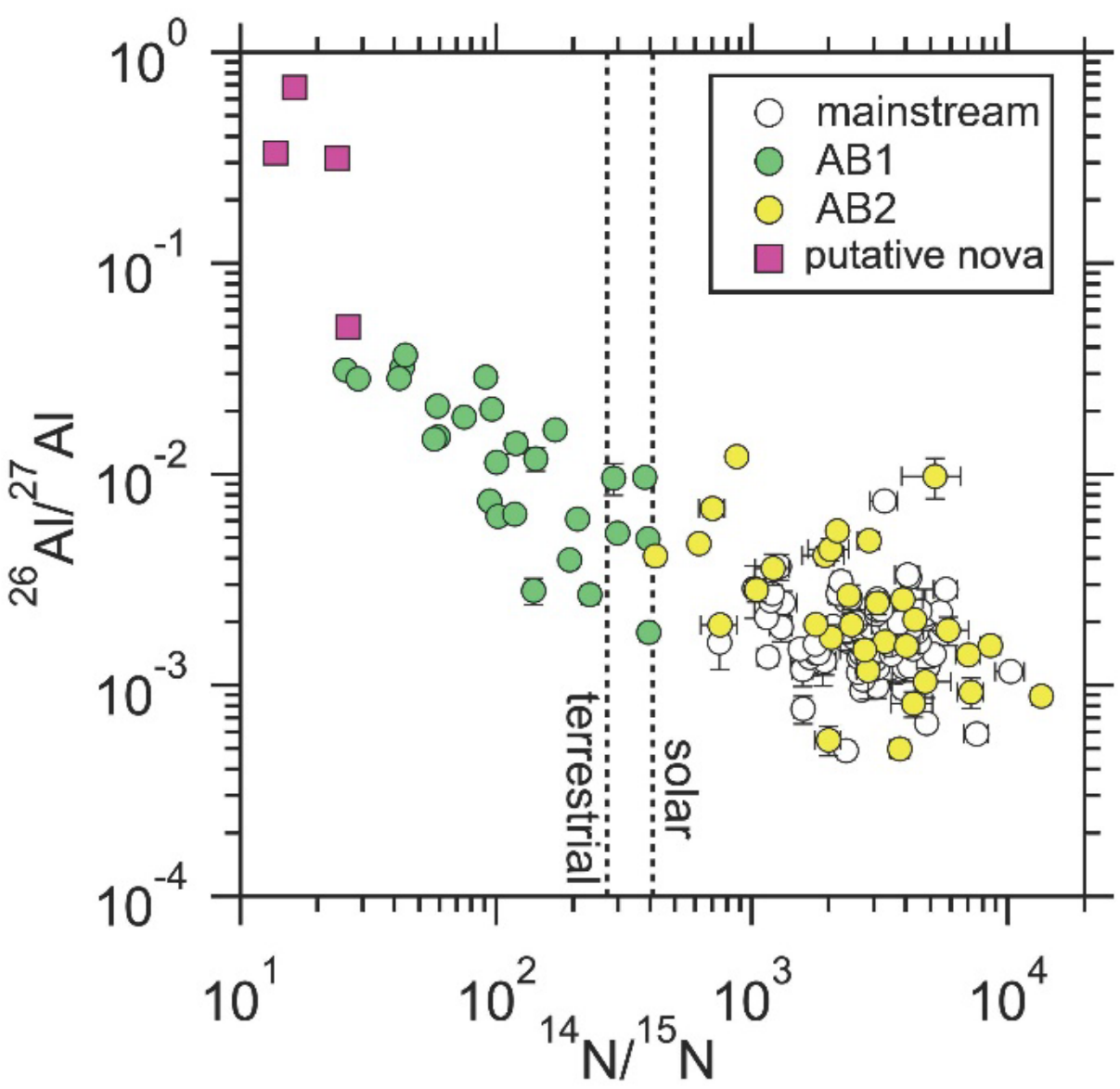

Initial $^{26}Al/^{27}Al$ ratios were inferred for 59 of the 79 AB grains and all the four putative nova grains, and are plotted along with ratios for 85 mainstream grains in in Fig. 3. Nineteen of the 20 AB grains without reported $^{26}Al/^{27}Al$ ratios belong to the subtype AB2. It is more challenging to determine low initial $^{26}Al/^{27}Al$ ratios as generally observed in AB2 grains because a lowered $^{26}Al/^{27}Al$ ratio means a smaller $^{26}Mg$ excess (with the Mg/Al ratio fixed), which can lead to a larger statistical uncertainty in the measured $\delta^{26}Mg$ value and, in some cases, no measurable $^{26}Mg$ excess within the uncertainty.

Our new inferred initial $^{26}Al/^{27}Al$ data lead to the following observations (Fig. 3). (*i*) AB2 grains overlap well with mainstream grains in their $^{14}N/^{15}N$ and $^{26}Al/^{27}Al$ ratios despite their distinct C isotope ratios. (*ii*) AB1 and AB2 grains together produce a positive trend between their $^{15}N$ and $^{26}Al$ enrichments. And (*iii*), the four putative nova grains deviate from the AB trend in Fig. 3 with more sharply increasing $^{26}Al$ excesses. The observation (*iii*) is in line with those observed in Section 3.2.1 that putative nova grains deviate systematically from AB grains in their C, N, and Si isotope ratios.

5.2.3 Titanium isotope data

**Fig. 4**. Titanium isotope ratios versus $\delta^{29}Si_{28}$ for grains from this study. The mainstream grain data are from [70]. The MSWD values shown are those for the respective linear fits to the mainstream grain data.

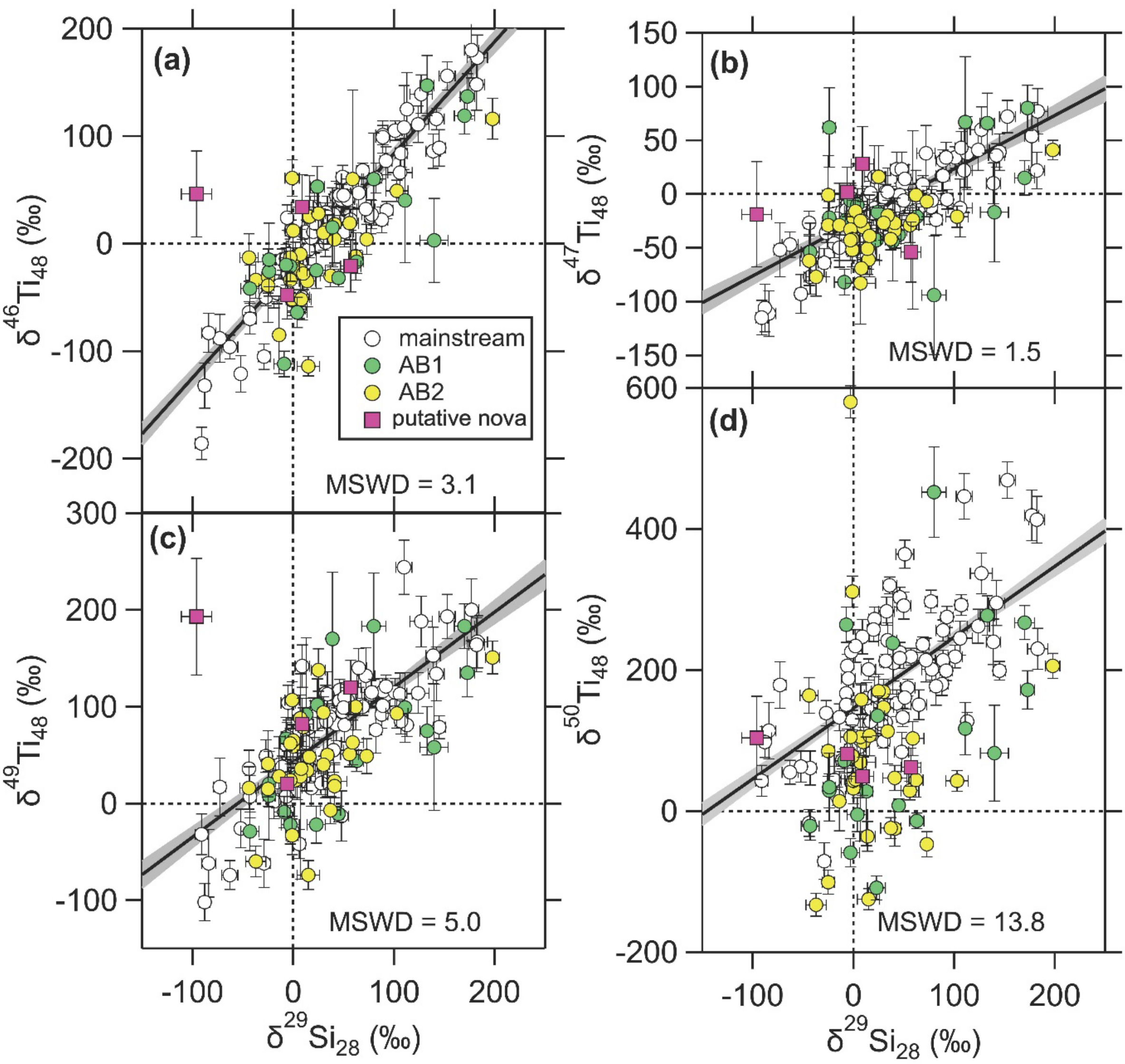

Titanium isotope data were obtained for 53 of the 79 AB grains, including 20 AB1 and 33 AB2 grains, and all four of the nova grains, and are compared to Ti isotope data for 76 mainstream grains in Fig. 4. The Ti isotope data presented here were obtained using a focused $O^-$ primary beam (~100 nm), providing significantly higher spatial resolution than in previous studies (~0.5–1 μm) [71] and allowing more reliable isolation of intrinsic grain signals from extrinsic contributions, as demonstrated in [77]. As discussed in Section 3.2.1, $\delta^{29}Si_{28}$ mainly reflects the effect of GCE among mainstream grains, which means that an isotope ratio would correlate with $\delta^{29}Si_{28}$ if it is also driven by GCE. In an AGB star, the *s*-process increases all $\delta^{i}Ti_{48}$ values by neutron capture over time, and the increases follow the order of $\Delta\delta^{50}Ti_{48}>\Delta\delta^{49}Ti_{48}>\Delta\delta^{46}Ti_{48}>\Delta\delta^{47}Ti_{48}$ [73, 78, 79]. As the MSWD value assess the goodness of a linear fit and the linear fits in Fig. 4 reflect the effects of GCE, the smaller the MSWD value, the larger the GCE effect, and the smaller the effect of AGB nucleosynthesis, and vice versa.

AB grains generally overlap with mainstream grains in their Ti isotope ratios, except for $\delta^{50}Ti_{48}$. Although AB1 and AB2 grains are plotted separately in Fig. 4, we discuss them together here because no statistically significant differences are resolved between the two subgroups in their Ti isotope compositions, given the limited number of grains and the relatively large analytical uncertainties. The MSWD values of the linear fits to AB grains follow the order observed among mainstream grains: $\delta^{50}Ti_{48}$ (MSWD = 33.1) > $\delta^{49}Ti_{48}$ (MSWD = 5.2) > $\delta^{46}Ti_{48}$ (MSWD = 4.5) > $\delta^{47}Ti_{48}$ (MSWD = 1.9). However, in the case of $\delta^{50}Ti_{48}$ the MSWD value is significantly larger for AB grains than for mainstream grains, reflecting a significantly wider range of neutron exposures experienced by the parent stars of AB grains (Fig. 4d). As shown in Fig. 4d, a significant fraction of the AB grains is shifted to lower $\delta^{50}Ti_{48}$ values (median value = 71‰) when compared to mainstream grains (median value = 199‰), pointing to neutron exposures that were weaker than the *s*-process in low-mass AGB stars. The weaker *s*-process effects observed in AB grains are unlikely to have been caused by higher initial stellar metallicities, i.e., lower neutron/seed ratios, given the lower $\delta^{29}Si_{28}$ values of AB grains on average, i.e., lower initial stellar metallicities for the parent stars of AB grains. It is, however, difficult to quantify neutron exposure based on $\delta^{50}Ti_{48}$. This is because mainstream grains tend to exhibit higher $\delta^{50}Ti_{48}$ values with increasing $\delta^{29}Si_{28}$ (Fig. 4d), pointing to the influence of GCE. The $\delta^{50}Ti_{48}$–$\delta^{29}Si_{28}$ correlation, however, is poor (MSWD = 13.8) with large scatter, likely caused by the significant production of $^{50}Ti$ by the *s*-process in their parent AGB stars. It is thus challenging to disentangle the effects of GCE and neutron capture on $\delta^{50}Ti_{48}$.

Regarding the putative nova grains, while three of the four grains overlap with the AB grains in Fig. 4, grain M5-A3-0913 lies significantly above the mainstream lines in Figs. 4a,c, which could be explained by either (*i*) production of $^{46,49}Ti$ (and maybe $^{50}Ti$) by neutron captures in its parent star, or (*ii*) destruction of $^{29}Si$ as predicted for the ejecta of massive CO novae and all ONe novae [80], leading to the lowered $\delta^{29}Si_{28}$ compared to the initial composition.

5.2.4 Nickel isotope data

Nickel isotope data were obtained for 28 of the 79 AB grains, including 12 AB1 and 16 AB2 grains, and are compared to data for 69 mainstream grains in Fig. 5. In contrast to $\delta^{50}Ti_{48}$, $\delta^{64}Ni_{58}$ shows no clear correlation with $\delta^{29}Si_{28}$ among mainstream grains (slope = $-0.95\pm0.62$), suggesting negligible influence from GCE effects on $\delta^{64}Ni_{58}$ in AGB stars. Consistent with this observation, Liu et al. [70] showed that the $\delta^{64}Ni_{58}$ values of mainstream grains increase with increases in the other Ni isotope ratios. This points to the coproduction of the minor Ni isotopes in AGB stars, which is well explained by the magnetic FRUITY AGB models that are calibrated against the heavy-element isotopic compositions of mainstream grains [81]. Thus, $\delta^{64}Ni_{58}$ can be used as a more reliable probe of neutron exposure than $\delta^{50}Ti_{48}$.

**Fig. 5**. Nickel isotope ratios versus $\delta^{29}Si_{28}$ for grains from this study. The mainstream grain data are from [70].

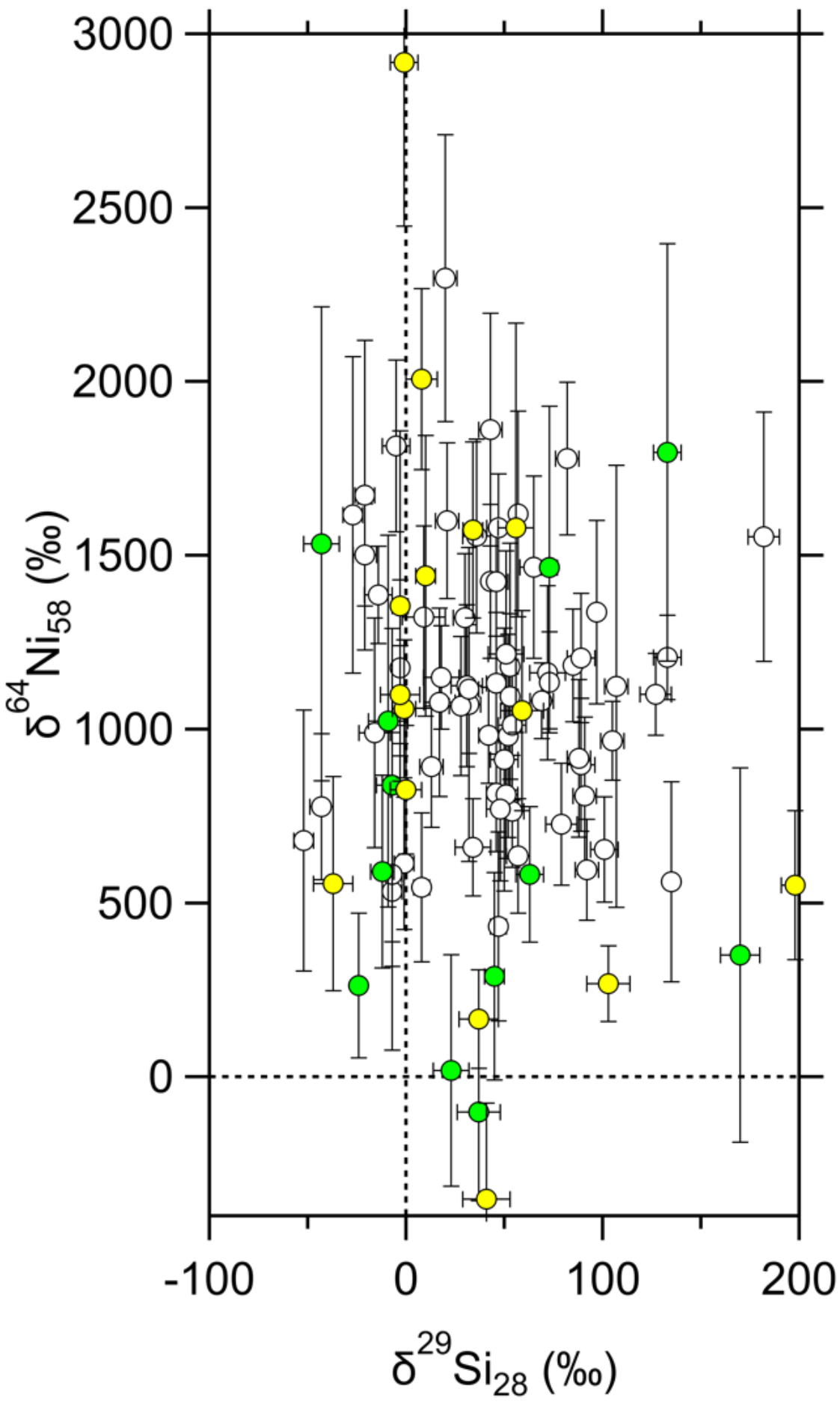

Compared to mainstream grains, AB grains show a wider range of $\delta^{64}Ni_{58}$ values, and a significant fraction of the grains (~25%), including both AB1 and AB2 grains, show smaller $^{64}Ni$ excesses (Fig. 5). This observation is consistent with the previous conclusion based on Ti isotope data that AB grains record a wider range of neutron exposures, which mainly resulted from reduced neutron exposures in some of their parent stars. Given the large $^{64}Ni$ excesses observed in some of the AB1 and AB2 grains, these AB grains clearly did not come from novae given the lack of neutron capture nucleosynthesis during nova explosions. We will thus focus on AB grains with small $\delta^{64}Ni_{58}$ and/or small $\delta^{50}Ti_{48}$ values (<50‰) (highlighted with * in Table 8) in the next Section to explore whether they could have originated in novae.

None of the putative nova grains in this study had sufficient Ni for deriving meaningful Ni isotope ratios. However, among the dozen putative nova grains that have been analyzed for their Ti isotope ratios in this and previous studies [31, 33, 34], all but two grains from [33] had close-to-solar $\delta^{50}Ti_{48}$ values ($-100‰<\delta^{50}Ti_{48}<150‰$), suggesting quite weak or no neutron capture nucleosynthesis in their parent stars. In our previous study of heavy-element isotope data collected using the CHILI instrument [47, 54, 76, 82], we also obtained Mo and Ba isotope data for three of the putative nova grains (M1-A8-G145; M2-A1-G410; M2-A4-G27) [31], that were all solar with 2σ errors. The solar heavy-element isotope data thus suggest the lack of neutron exposure in the parent stars of putative nova grains, although we cannot exclude the possibility of severe Mo and Ba contamination during the CHILI analyses. In summary, in contrast to the AB grains, none of the putative nova grains investigated so far have shown any signatures of neutron capture, in line with an origin in novae. Thus, the four putative nova grains will be included for discussion in the next Section.

**6. Comparison Between Grain Data and Nova Nucleosynthesis Predictions**

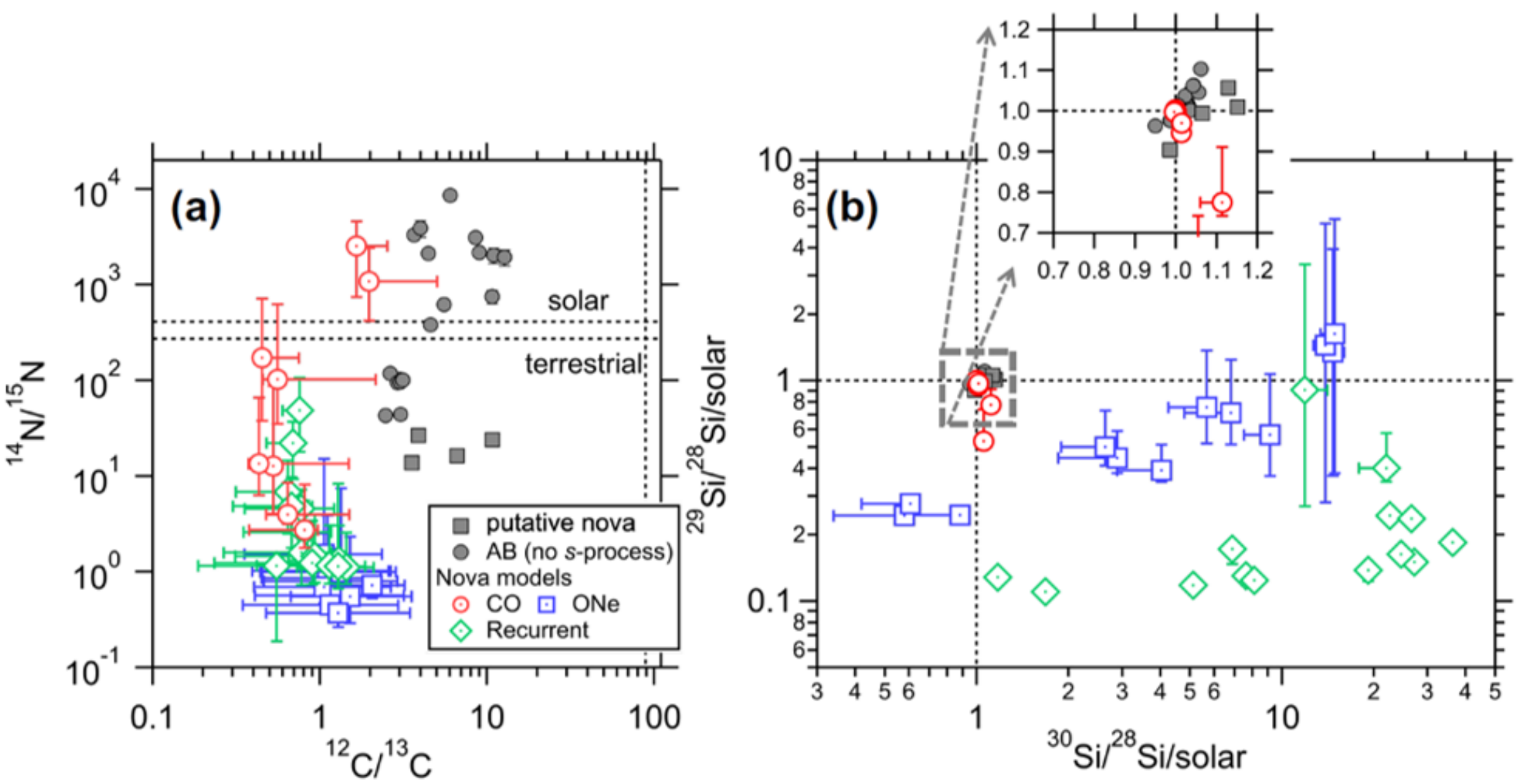

**Figure 6.** Carbon, N, and Si isotopic ratios. Nova grains and selected AB grains without obvious *s*-process enrichments ($\delta^{50}Ti_{48}$ <50‰ and without any significant $^{60}Ni$ excesses; highlighted with * in Table 8) are shown alongside nova model predictions. Model symbols denote mean ejecta compositions, with bars indicating shell-to-shell variations. In panel (b), absolute isotopic ratios, normalized to solar values, are plotted rather than δ-values to aid visualization. The inset (linear scale) shows an expanded view of the tightly clustered CO nova models and grain data within the region outlined by the grey dashed square.

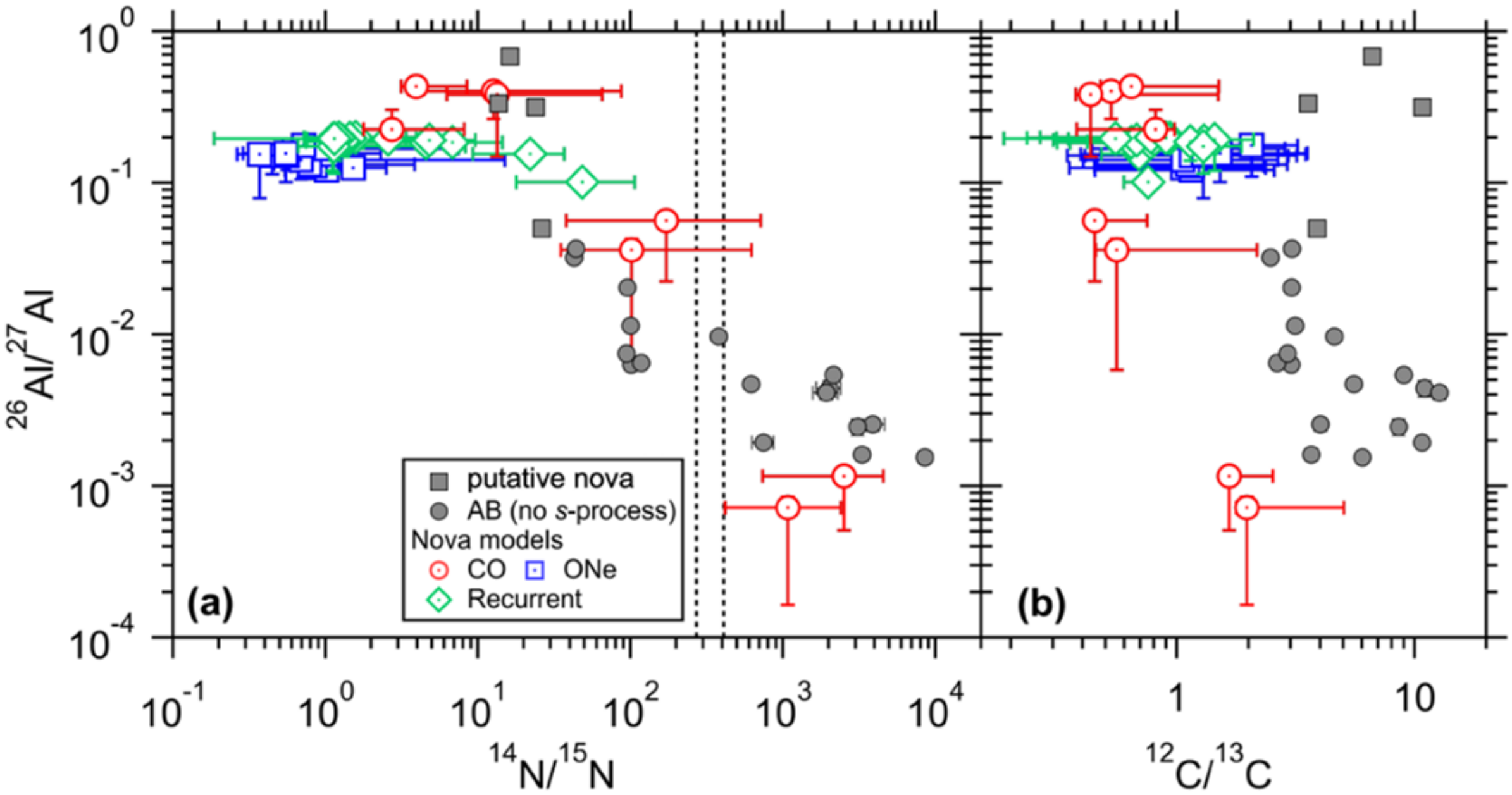

**Figure 7.** Plots of $^{26}Al/^{27}Al$ versus $^{14}N/^{15}N$ and $^{12}C/^{13}C$. The same set of grains as in Fig. 6 are plotted for comparison with the same nova models.

Grain data and nova model predictions for a subset of isotopic ratios are shown in Figs. 6 and 7. A clear and consistent pattern emerges when all isotope systems are considered together: CO nova models provide the best overall agreement with the combined C, N, Al-Mg, Si, and Ti isotopic compositions of the $^{13}C$-rich SiC grains, whereas ONe and recurrent nova models fail to simultaneously reproduce the key isotopic constraints. In the following, we evaluate the compatibility of each nova type with the grain data and discuss the implications for the stellar origins of AB and putative nova grains.

### 6.1. Consistency between CO Nova Models and Grain Data

CO nova models offer the closest overall match to the isotopic signatures of the majority of $^{13}$C-rich SiC grains. Although all nova models, including CO novae, overproduce $^{13}$C and therefore predict lower-than-observed $^{12}$C/$^{13}$C ratios, CO novae best reproduce the full set of isotopic systems when all constraints are considered simultaneously. The CO models span the range of $^{14}$N/$^{15}$N ratios observed in both AB and nova grains and, more importantly, closely reproduce the trend of decreasing $^{14}$N/$^{15}$N with increasing $^{26}$Al/$^{27}$Al defined by the putative nova and AB grains in Fig. 7a. Their Si isotopic predictions also align with the patterns discussed in Section 5. Low- to intermediate-mass CO nova models (0.6–1.0 $M_{\odot}$) generally preserve the initial solar $\delta^{29}Si_{28}$ and $\delta^{30}Si_{28}$ values, consistent with the observation that the Si isotopic compositions of AB grains are dominated by GCE (Section 5). Notably, the 1.0 $M_{\odot}$ CO models (CO5 and CO6) predict, compared to the initial solar composition, slight depletions in $^{29}$Si and mild enrichments in $^{30}$Si, providing a natural explanation for the small but systematic displacement of putative nova grains toward higher $\delta^{30}Si_{28}$ relative to the AB population in Fig. 6b. These trends imply that the Si isotopic compositions of AB grains primarily reflect their initial stellar compositions—a conclusion independently corroborated by the mainstream SiC data presented in Section 5—whereas the nova grains record mild but measurable Si-isotope processing in CO novae. Titanium isotopes further reinforce this conclusion: both AB and nova grains show $\delta^{46}Ti_{48}$ and $\delta^{47}Ti_{48}$ values that follow the mainstream GCE trend (Fig. 4), and CO nova models predict no effects on Ti isotopes, consistent with the data. Taken together, these constraints indicate that CO novae—the most common type of Galactic nova—provide a self-consistent and astrophysically plausible explanation for the majority of AB and putative nova grains.

ONe nova models show more limited agreement with the grain data. Low-mass ONe models (1.15–1.25 $M_{\odot}$) can account for the two extreme Si isotopic patterns observed in the putative nova grains of Nittler and Hoppe [33]: large, coupled $\delta^{29}Si_{28}$–$\delta^{30}Si_{28}$ deficits and combinations of $\delta^{29}Si_{28}$ deficits with pronounced $\delta^{30}Si_{28}$ excesses. However, these models simultaneously predict $^{12}$C/$^{13}$C ratios that are too low, $^{14}$N/$^{15}$N ratios far below the grain values, and $^{26}$Al/$^{27}$Al ratios (~0.1–0.15) significantly below those inferred for the same grains. Importantly, the published $^{26}$Al/$^{27}$Al ratios of 0.27 and 0.39 from Nittler and Hoppe [33] are now known to be underestimates because the Mg/Al relative sensitivity factor used in earlier NanoSIMS studies underestimated the initial ratios by a factor of two [62]. Since the dilution needed to reconcile the C, N, and Si isotopes would further lower the predicted $^{26}$Al/$^{27}$Al ratios, the discrepancy between ONe nova predictions and the grain data only increases. Thus, although ONe novae can potentially explain some of the most extreme Si isotopic signatures reported in the literature, they fail to reproduce the full multielement isotopic patterns and cannot consistently account for any of the $^{13}$C-rich grains considered here.

Recurrent nova models are inconsistent with the grain data across all isotopic systems. Their $\delta^{29}Si_{28}$ and $\delta^{30}Si_{28}$ values do not overlap with any AB or nova grains, either from this study or from previous work. The models predict extremely negative $\delta^{29}Si_{28}$ values (−900 to −600‰) and large $\delta^{30}Si_{28}$ enrichments ranging from hundreds to tens of thousands of permil, entirely unlike the modest Si anomalies observed in the grains. Some recurrent nova models also produce exceptionally large $\delta^{46}Ti_{48}$ enrichments (up to ~$10^3$–$10^5$‰), whereas AB and putative nova grains exhibit $\delta^{46}Ti_{48}$ and $\delta^{47}Ti_{48}$ ratios dominated by GCE. Even the Ti-anomalous putative nova grain reported by Nittler and Hoppe [33], which shows a $^{47}$Ti excess but solar $\delta^{46}Ti_{48}$, disagrees with recurrent-nova predictions. These combined C, N, Al-Mg, Si, and Ti isotopic mismatches rule out recurrent novae as viable sources for any known $^{13}$C-rich SiC grains.

In summary, the combined isotopic constraints indicate that CO novae provide the most coherent explanation for the majority of AB grains (i.e., those without clear isotopic signatures of the *s*-process) and putative nova grains. The model predictions align with the grain data across all major isotopic systems and are consistent with the expectation that CO novae dominate the Galactic nova population. ONe nova models can reproduce only the most extreme Si anomalies reported in earlier studies but fail to match the accompanying C, N, and Al-Mg isotopic signatures simultaneously. Recurrent novae are incompatible with the grain data in every isotopic system. These results collectively support the conclusion that CO novae are the primary contributors to the population of $^{13}$C-rich presolar SiC grains, with only rare contributions from low-mass ONe novae to the most Si-anomalous grains.

### 6.2. Comparison with Previous Work

These conclusions refine and substantially extend earlier analyses by Iliadis et al. [35] and Bose & Starrfield [30], who reexamined previously published isotopic data for putative nova grains. Because only C, N, and Si isotope ratios were available for most literature grains, their interpretations necessarily relied on a restricted subset of isotopic constraints and could not incorporate the multielement information—particularly Mg-Al and Ti isotopes—used in this study. In addition, the literature $^{26}$Al/$^{27}$Al values considered in those studies span a broad and generally lower range (~0.01–0.4) than the ratios measured here. As discussed earlier, these discrepancies likely reflect analytical limitations in the older datasets, including less effective suppression of Al contamination and use of an Mg/Al relative sensitivity factor now known to underestimate $^{26}$Al/$^{27}$Al in SiC by a factor of two.

Building on these literature data, Bose & Starrfield [30] used a suite of CO nova models spanning white dwarf masses from 0.6 to 1.35 $M_\odot$ and concluded that models of 0.8–1.35 $M_\odot$ (pre-enriched by 25–50%) provided the closest matches to the previously published putative nova grain compositions. However, white dwarfs $\gtrapprox$1.1 $M_\odot$ are generally expected to be ONe-rich rather than CO-rich, e.g., [83-85], meaning that part of their model grid may not correspond to astrophysically realistic novae. Moreover, because most of the grains examined in [30, 35] lacked Ti and Ni isotope measurements, their interpretations necessarily relied on a more restricted set of isotopic diagnostics than is available in this study.

By contrast, the NanoSIMS imaging analyses used in this study—together with the revised Mg/Al RSF—provide the most accurate multielement characterization to date of grains exhibiting nova-like $^{13}$C enrichments. When paired with the expanded grid of hydrodynamic nova models developed here, these improved data reveal a far more coherent and self-consistent agreement with CO nova nucleosynthesis than was possible in previous work. The inclusion of AB grains, which were not considered in [30, 35], further strengthens this conclusion: the isotopic differences between the subset of AB grains lacking *s*-process signatures and grains traditionally interpreted as nova-originated, i.e., putative nova grains, can be reproduced by varying white dwarf mass and pre-enrichment parameters in the CO nova models, indicating that CO novae contribute to both populations.

### 6.3. Remaining Issues

A remaining challenge is the persistent discrepancy in $^{12}$C/$^{13}$C (up to a factor ~10), which all nova models in this study underpredict relative to the grain data. A recent Monte Carlo sensitivity study by [86] showed that thermonuclear reaction-rate uncertainties generally affect $^{12}C/^{13}C$ and several other key isotopic ratios at the < 20% level, insufficient to resolve the mismatch. This suggests that the discrepancy reflects limitations in current nova models rather than nuclear physics uncertainties. In particular, mixing at the white dwarf–envelope interface remains poorly constrained in 1D simulations; multidimensional processes such as shear, convective boundary mixing, and diffusion likely play important roles. Addressing the $^{12}C/^{13}C$ mismatch will therefore require both improved physical treatments of mixing and a broader exploration of the nova parameter space.

A related issue concerns the conditions under which SiC dust can form in nova ejecta. Most of the CO nova models presented here predict bulk, mass-averaged ejecta compositions that are O-rich, seemingly disfavoring SiC condensation. However, extensive infrared and ultraviolet observations demonstrate that dust formation is a common outcome of classical nova eruptions, typically occurring months after optical maximum, at the time when a pronounced optical decline and enhanced infrared emission is detected [87-91]. CO novae, in particular, are observed to be prolific dust producers, whereas ONe novae exhibit dust-formation episodes far less frequently. The lower densities and higher expansion velocities characteristic of ONe-nova ejecta may inhibit efficient grain condensation, further favoring CO novae as plausible sources of putative nova and AB SiC grains.

Observations suggest that nova explosions produce rich mineralogies. However, the coexistence of O-rich and C-rich dust observed in some novae [91] appears difficult to reconcile with classical equilibrium condensation arguments [32]. Most heteronuclear diatomic molecules have dissociation energies of approximately 3–5 eV; however, carbon monoxide (CO) has a much higher bond energy of 11.2 eV and can therefore be dissociated only by high-energy photons. In the absence of intense, high-energy radiation fields, condensation in an O-rich plasma (O > C) results in nearly all C atoms being locked into very stable CO molecules. Consequently, the formation of C-rich minerals is not expected. Conversely, in a C-rich environment (C > O), all O is consumed in CO formation and, in principle, oxidized condensates cannot form.

When individual ejected shells are examined instead of mass-averaged compositions, substantial compositional gradients emerge. In four of the eight CO nova models discussed here, specific shells attain C/O > 1, while two additional models approach unity in localized regions. Such shell-to-shell variations naturally permit the coexistence of C-rich and O-rich condensation environments within a single nova event, without requiring globally C-rich ejecta[6]. This interpretation is strongly supported by the presolar grain evidence: both putative nova grains and AB SiC grains exhibit well-crystallized cubic structures with relatively low defect densities [69], closely resembling mainstream SiC grains formed around low-mass AGB stars. These microstructural characteristics argue for relatively quiescent condensation environments, rather than highly radiation-dominated or kinetically extreme conditions. The formation of SiC in localized, C-rich pockets within an otherwise O-rich nova outflow therefore provides a physically plausible pathway that is consistent with both hydrodynamic nova models and grain crystallography.

It is worth noting, however, that the physical conditions that characterize the nova ejecta are substantially different than those in AGB stars, with plasma moving at several thousand km/s and irradiated by an intense radiation field by the underlying white dwarf. In these conditions, it has been proposed that dust condensation in novae may proceed under nonequilibrium, kinetic conditions, driven by induced dipole reactions in the ejecta, rather than at equilibrium [92].

[6]Equilibrium condensation studies have also revealed that the presence of large amounts of intermediate-mass elements, such as Al, Ca, Mg, and Si, may alter the condensation pathways, enabling the formation of C-rich dust even in environments where O > C [59].

The large $^{15}$N enrichments characteristic of putative nova and AB1 grains place strong constraints on their stellar origins. Such signatures require operation of the hot CNO cycle, which is currently predicted to occur only in explosive environments [31]. This restriction effectively limits viable sources to CCSNe and novae. As shown above, low- to intermediate-mass CO novae provide a natural explanation for the observed N–Al isotopic systematics of putative nova and AB1 grains, including the pronounced $^{14}$N/$^{15}$N–$^{26}$Al/$^{27}$Al trend and the absence of neutron-capture signatures in a subset of the grains. Within this framework, CO novae emerge as a well-motivated and astrophysically plausible contributor to the $^{13}$C-,$^{15}$N-rich grain population, even though remaining discrepancies in $^{12}$C/$^{13}$C point to the need for further refinement of nova models rather than to a failure of the nova scenario itself. For example, the initial composition adopted in the CO nova models presented in this study assumes solar abundances for the material accreted from the companion star, mixed with material from the outermost layers of the white dwarf. The latter is characterized by mass fractions of X($^{12}$C) = X($^{16}$O) = 0.495 and X($^{22}$Ne) = 0.01, values intended to be representative of the composition of these layers. Modest variations in the initial $^{12}$C and $^{16}$O content may affect the final C/O ratios in the ejecta and, in turn, alter the resulting $^{12}$C/$^{13}$C ratios.

The stellar origins of AB2 grains ($^{14}$N-rich) remain more ambiguous. Their nitrogen isotopic compositions reflect cold CNO-cycle processing, which operates during H-burning in a wide range of stellar environments, including born-again AGB stars, J-type carbon stars, and potentially supernova progenitors [29, 53, 93]. Our results show that AB2 grains lacking *s*-process signatures can also be accommodated by the low-mass CO nova models, extending the possible nova contribution beyond putative nova and AB1 grains. However, AB2 grains that exhibit *s*-process enrichments comparable to or exceeding those of mainstream grains require additional stellar sources, most plausibly born-again AGB stars. Born-again AGB stars are post-AGB objects that undergo a late thermal pulse after leaving the AGB, potentially leading to unusual surface compositions. Existing born-again AGB nucleosynthesis models have focused on reproducing chemically peculiar objects such as Sakurai's Object via the intermediate neutron-capture process (*i*-process) [94], which can explain the exotic Ca–Ti isotopic signatures observed in some high-density presolar graphite grains [95] but are incompatible with the isotopic signatures of AB2 SiC grains [47]. Given the large and largely unexplored parameter space of born-again AGB evolution, future modeling efforts will be essential to assess whether alternative born-again conditions—without strong *i*-process activation—can account for the *s*-process–rich subset of AB2 grains.

## 7. Conclusions

Our results demonstrate that low- to intermediate-mass (0.6–1.0 $M_\odot$) CO novae provide the most coherent and astrophysically plausible explanation for the majority of $^{13}$C-rich presolar SiC grains. When the full suite of isotopic systems—C, N, Mg–Al, Si, Ti, and Ni—is considered simultaneously, CO nova models consistently yield the closest match to both the four newly identified putative nova grains and to a subset of AB grains lacking *s*-process signatures. This level of agreement reflects two key advances of the present study: (*i*) high-fidelity multielement NanoSIMS data enabled by high-resolution imaging and the revised

Mg/Al RSF for SiC, and (*ii*) an expanded grid of hydrodynamic nova models that permits controlled exploration of white-dwarf mass and pre-enrichment. Together, these developments reveal a unified and self-consistent nova interpretation that was not accessible in earlier work relying on limited isotopic constraints and older datasets.

The combined grain-model comparisons clarify the distinct roles of different nova subclasses. Low- to intermediate-mass CO novae (0.6–0.8 $M_\odot$) largely preserve the initial Si isotopic composition inherited from the interstellar medium, in agreement with the observation that AB grains follow the Si- and Ti-isotope GCE trends. At the same time, slightly higher-mass CO models (~1.0 $M_\odot$) naturally account for the mild but measurable Si-isotope shifts observed in the nova grains. These CO novae also span the full range of $^{14}$N/$^{15}$N ratios displayed by the grains and reproduce the strong $^{14}$N/$^{15}$N–$^{26}$Al/$^{27}$Al anticorrelation that characterizes our dataset. In contrast, ONe novae match only extreme Si anomalies reported in earlier studies and cannot reproduce the accompanying C, N, and Mg–Al isotopic signatures without invoking large dilution, while recurrent novae are incompatible with the measured isotopes across all systems. Taken together, these distinctions underscore the dominant contribution of CO novae to the $^{13}$C-rich SiC population, with only limited roles for other nova types.

**Acknowledgements**

This work has been partially supported by the Spanish MINECO grant PID2023-148661NB-I00, and by the E.U. FEDER funds. N.L. acknowledges support from NASA through grants 80NSSC23K1034 and 80NSSC24K0132. This paper is dedicated to the memory and legacy of Roberto Gallino. We especially honor his visionary role in initiating and nurturing a number of fruitful collaborations between cosmochemists and stellar modelers. In particular, he instigated a collaboration between the cosmochemistry group at Washington University in St. Louis and the astrophysics group at the Technical University of Catalonia in Barcelona, at the pivotal moment when the first presolar grains of putative nova origin were identified. The ongoing collaborations among diverse research groups in this field stand as a lasting testament to his scientific vision and leadership, which continue inspiring our community today.

**References**

1. S. Starrfield, C. Iliadis, and W.R. Hix, Thermonuclear processes, in *Classical Novae* (Ed. M.F. Bode and A. Evans (Exec., 77-101 (2008) https://doi.org/10.1017/cbo9780511536168.006.
2. S. Starrfield, C. Iliadis, and W.R. Hix, The thermonuclear runaway and the classical nova outburst. *Publ. Astron. Soc. Pac.* **128**, 051001 (2016) https://doi.org/10.1088/1538-3873/128/963/051001.
3. J. José and S.N. Shore, Observational mysteries and theoretical challenges for abundance studies, in *Classical Novae* (Ed. M.F. Bode and A. Evans (Exec., 121-149 (2008) https://doi.org/10.1017/cbo9780511536168.008.
4. J. José, *Stellar explosions: Hydrodynamics and nucleosynthesis*. Boca Raton: CRC Press. (2016) https://doi.org/10.1201/b19165.
5. L. Chomiuk, B.D. Metzger, and K.J. Shen, New insights into classical novae. *Annu. Rev. Astron. Astrophys.* **59**, 391-444 (2021) https://doi.org/10.1146/annurev-astro-112420-114502.
6. A.W. Shafter, The Galactic nova rate revisited. *Astrophys. J.* **834**, 196 (2017) https://doi.org/10.3847/1538-4357/834/2/196.
7. S.W. Jha, Type Iax supernovae, in *Handbook of Supernovae* (Ed. A.W. Alsabti and P. Murdin (Exec., 375 (2017) https://doi.org/10.1007/978-3-319-21846-5_42.
8. J. José and M. Hernanz, Hydrodynamic simulations of the recurrent nova T Coronae Borealis: Nucleosynthesis predictions. *Astron. Astrophys.* **698**, A251 (2025)

https://doi.org/10.1051/0004-6361/202553762.

9. S. Starrfield, M. Bose, C.E. Woodward, et al., Hydrodynamic predictions for the next outburst of T Coronae Borealis: It will be the brightest classical or recurrent nova ever observed in X-rays. *Astrophys. J.* **982**, 89 (2025) https://doi.org/10.3847/1538-4357/adb8ed.
10. W.M. Sparks, S. Starrfield, and J.W. Truran, A hydrodynamic study of a slow nova outburst. *Astrophys. J.* **220**, 1063-1075 (1978) https://doi.org/10.1086/155992.
11. D. Prialnik, M.M. Shara, and G. Shaviv, The evolution of a slow nova model with a Z = 0.03 envelope from pre-explosion to extinction. *Astron. Astrophys.* **62**, 339-348 (1978)
12. S. Starrfield, J.W. Truran, W.M. Sparks, and G.S. Kutter, CNO abundances and hydrodynamic models of the nova outburst. *Astrophys. J.* **176**, 169 (1972) https://doi.org/10.1086/151619.
13. S. Starrfield, J.W. Truran, and W.M. Sparks, CNO abundances and hydrodynamic studies of the nova outburst. V. 1.00 M sun models with small mass envelopes. *Astrophys. J.* **226**, 186-202 (1978) https://doi.org/10.1086/156598.
14. M. Livio and J.W. Truran, Elemental mixing in classical nova systems. *Annals of the New York Academy of Sciences* **617**, 126-137 (1990) https://doi.org/10.1111/j.1749-6632.1990.tb37801.x.
15. R. Rosner, A. Alexakis, Y.-N. Young, J.W. Truran, and W. Hillebrandt, On the C/O enrichment of nova ejecta. *Astrophys. J.* **562**, L177-L179 (2001) https://doi.org/10.1086/338327.
16. A. Alexakis, A.C. Calder, A. Heger, et al., On heavy element enrichment in classical novae. *Astrophys. J.* **602**, 931-937 (2004) https://doi.org/10.1086/381086.
17. J. Casanova, J. José, E. García-Berro, S.N. Shore, and A.C. Calder, Kelvin-Helmholtz instabilities as the source of inhomogeneous mixing in nova explosions. *Nature* **478**, 490-492 (2011) https://doi.org/10.1038/nature10520.
18. J. Casanova, J. José, E. García-Berro, and S.N. Shore, Three-dimensional simulations of turbulent convective mixing in ONe and CO classical nova explosions. *Astron. Astrophys.* **595**, A28 (2016) https://doi.org/10.1051/0004-6361/201628707.
19. J. José, S.N. Shore, and J. Casanova, 123-321 models of classical novae. *Astron. Astrophys.* **634**, A5 (2020) https://doi.org/10.1051/0004-6361/201936893.
20. A. Evans, Y.V. Pavlenko, D.P.K. Banerjee, et al., Gas phase SiO in the circumstellar environment of the recurrent nova T Coronae Borealis. *Mon. Not. R. Astron. Soc.* **486**, 3498-3505 (2019) https://doi.org/10.1093/mnras/stz1071.
21. A. Evans, D.P.K. Banerjee, T.R. Geballe, et al., Near-infrared spectroscopy of the LMC recurrent nova LMCN 1968-12a. *Mon. Not. R. Astron. Soc.* **536**, 1710-1717 (2025) https://doi.org/10.1093/mnras/stae2711.
22. Y.V. Pavlenko, A. Evans, D.P.K. Banerjee, et al., Isotopic ratios in the red giant component of the recurrent nova T Coronae Borealis. *Mon. Not. R. Astron. Soc.* **498**, 4853-4863 (2020) https://doi.org/10.1093/mnras/staa2658.
23. N. Liu, Presolar grains, in *Treatiese on Geochemistry (Third Edition)* (Exec. Elsevier (2024) https://doi.org/https://arxiv.org/abs/2406.14694.
24. P.R. Heck, J. Greer, L. Kööp, et al., Lifetimes of interstellar dust from cosmic ray exposure ages of presolar silicon carbide. *Proceedings of the National Academy of Science* **117**, 1884-1889 (2020) https://doi.org/10.1073/pnas.1904573117.
25. S. Amari, X. Gao, L.R. Nittler, et al., Presolar grains from novae. *Astrophys. J.* **551**, 1065-1072 (2001) https://doi.org/10.1086/320235.
26. F. Gyngard, E. Zinner, L.R. Nittler, et al., Automated NanoSIMS measurements of spinel stardust from the Murray meteorite. *Astrophys. J.* **717**, 107-120 (2010) https://doi.org/10.1088/0004-637x/717/1/107.
27. A.N. Nguyen and S. Messenger, Resolving the stellar sources of isotopically rare presolar silicate grains through Mg and Fe isotopic analyses. *Astrophys. J.* **784**, 149 (2014) https://doi.org/10.1088/0004-637x/784/2/149.
28. P. Haenecour, J.Y. Howe, T.J. Zega, et al., Laboratory evidence for co-condensed oxygen- and carbon-rich meteoritic stardust from nova outbursts. *Nat. Astron.* **3**, 626-630 (2019) https://doi.org/10.1038/s41550-019-0757-4.
29. S. Amari, L.R. Nittler, E. Zinner, K. Lodders, and R.S. Lewis, Presolar SiC grains of type A and B: their isotopic compositions and stellar origins. *Astrophys. J.* **559**, 463-483 (2001)

https://doi.org/10.1086/322397.
30. M. Bose and S. Starrfield, Condensation of SiC stardust in CO nova outbursts. *Astrophys. J.* **873**, 14 (2019) https://doi.org/10.3847/1538-4357/aafc2f.
31. N. Liu, L.R. Nittler, M. Pignatari, C.M.O'D. Alexander, and J. Wang, Stellar origin of $^{15}$N-rich presolar SiC grains of type AB: supernovae with explosive hydrogen burning. *Astrophys. J. Lett.* **842**, L1 (2017) https://doi.org/10.3847/2041-8213/aa74e5.
32. K. Lodders and B. Fegley, Jr., The origin of circumstellar silicon carbide grains found in meteorites. *Meteoritics* **30**, 661 (1995) https://doi.org/10.1111/j.1945-5100.1995.tb01164.x.
33. L.R. Nittler and P. Hoppe, Are presolar silicon carbide grains from novae actually from supernovae? *Astrophys. J. Lett.* **631**, L89-L92 (2005) https://doi.org/10.1086/497029.
34. N. Liu, L.R. Nittler, C.M.O'D. Alexander, et al., Stellar origins of extremely $^{13}$C- and $^{15}$N-enriched presolar SiC grains: novae or supernovae? *Astrophys. J.* **820**, 140 (2016) https://doi.org/10.3847/0004-637x/820/2/140.
35. C. Iliadis, L.N. Downen, J. José, L.R. Nittler, and S. Starrfield, On presolar stardust grains from CO classical novae. *Astrophys. J.* **855**, 76 (2018) https://doi.org/10.3847/1538-4357/aaabb6.
36. J.G. Barzyk, M.R. Savina, A.M. Davis, et al., Constraining the $^{13}$C neutron source in AGB stars through isotopic analysis of trace elements in presolar SiC. *Meteorit. Planet. Sci.* **42**, 1103-1119 (2007) https://doi.org/10.1111/j.1945-5100.2007.tb00563.x.
37. L.R. Nittler, C.M.O'D. Alexander, R. Gallino, et al., Aluminum-, calcium- and titanium-rich oxide stardust in ordinary chondrite meteorites. *Astrophys. J.* **682**, 1450-1478 (2008) https://doi.org/10.1086/589430.
38. S. Amari, E. Zinner, and R. Gallino, Presolar graphite from the Murchison meteorite: an isotopic study. *Geochim. Cosmochim. Acta* **133**, 479-522 (2014) https://doi.org/10.1016/j.gca.2014.01.006.
39. C. Floss and P. Haenecour, Presolar silicate grains: abundances, isotopic and elemental compositions, and the effects of secondary processing. *Geochemical Journal* **50**, 3-25 (2016) https://doi.org/10.2343/geochemj.2.0377.
40. N. Liu, J. Barosch, L.R. Nittler, et al., New multielement isotopic compositions of presolar SiC grains: implications for their stellar origins. *Astrophys. J. Lett.* **920**, L26 (2021) https://doi.org/10.3847/2041-8213/ac260b.
41. A. Boujibar, S. Howell, S. Zhang, et al., Cluster analysis of presolar silicon carbide grains: evaluation of their classification and astrophysical implications. *Astrophys. J. Lett.* **907**, L39 (2021) https://doi.org/10.3847/2041-8213/abd102.
42. G. Hystad, A. Boujibar, N. Liu, L.R. Nittler, and R.M. Hazen, Evaluation of the classification of pre-solar silicon carbide grains using consensus clustering with resampling methods: An assessment of the confidence of grain assignments. *Mon. Not. R. Astron. Soc.* **510**, 334-350 (2022) https://doi.org/10.1093/mnras/stab3478.
43. J. José, M. Hernanz, and C. Iliadis, Nucleosynthesis in classical novae. *Nucl. Phys. A* **777**, 550-578 (2006) https://doi.org/10.1016/j.nuclphysa.2005.02.121.
44. M. Hernanz, Gamma-rays from classical novae, in *Classical Novae* (Ed. M.F. Bode and A. Evans (Exec., 252-284 (2008) https://doi.org/10.1017/cbo9780511536168.013.
45. J. José, M. Hernanz, and A. Coc, New results on $^{26}$Al production in classical novae. *Astrophys. J.* **479**, L55-L58 (1997) https://doi.org/10.1086/310575.
46. J. José and M. Hernanz, Nucleosynthesis in classical novae: CO versus ONe white dwarfs. *Astrophys. J.* **494**, 680-690 (1998) https://doi.org/10.1086/305244.
47. N. Liu, T. Stephan, P. Boehnke, et al., J-type carbon stars: a dominant source of $^{14}$N-rich presolar SiC grains of type AB. *Astrophys. J. Lett.* **844**, L12 (2017) https://doi.org/10.3847/2041-8213/aa7d4c.
48. S. Amari, A. Anders, A. Virag, and E. Zinner, Interstellar graphite in meteorites. *Nature* **345**, 238-240 (1990) https://doi.org/10.1038/345238a0.
49. D.D. Clayton, $^{22}$Na, Ne-E, extinct radioactive anomalies and unsupported $^{40}$Ar. *Nature* **257**, 36-37 (1975) https://doi.org/10.1038/257036b0.
50. S. Amari, On the origin of 22Na in Ne-E(L). *Meteoritics and Planetary Science Supplement* **43**, 5271 (2008)
51. M. Pignatari, S. Amari, P. Hoppe, et al., Production of radioactive $^{22}$Na in core-collapse

supernovae: The Ne-E(L) component in presolar grains and its possible consequences on supernova observations. *Astrophys. J.* **990**, 19 (2025) https://doi.org/10.3847/1538-4357/adef4c.

52. T. Rauscher, A. Heger, R.D. Hoffman, and S.E. Woosley, Nucleosynthesis in massive stars with improved nuclear and stellar physics. *Astrophys. J.* **576**, 323-348 (2002) https://doi.org/10.1086/341728.
53. M. Pignatari, E. Zinner, P. Hoppe, et al., Carbon-rich presolar grains from massive stars: subsolar $^{12}C/^{13}C$ and $^{14}N/^{15}N$ Ratios and the Mystery of $^{15}N$. *Astrophys. J.* **808**, L43 (2015) https://doi.org/10.1088/2041-8205/808/2/l43.
54. N. Liu, T. Stephan, P. Boehnke, et al., Common occurrence of explosive hydrogen burning in Type II supernovae. *Astrophys. J.* **855**, 144 (2018) https://doi.org/10.3847/1538-4357/aaab4e.
55. P. Hoppe, M. Pignatari, J. Kodolányi, E. Gröner, and S. Amari, NanoSIMS isotope studies of rare types of presolar silicon carbide grains from the Murchison meteorite: Implications for supernova models and the role of $^{14}C$. *Geochim. Cosmochim. Acta* **221**, 182-199 (2018) https://doi.org/10.1016/j.gca.2017.01.051.
56. A.L. Sallaska, C. Iliadis, A.E. Champange, et al., STARLIB: A next-generation reaction-rate library for nuclear astrophysics. *Astrophys. J. Suppl. Ser.* **207**, 18 (2013) https://doi.org/10.1088/0067-0049/207/1/18.
57. K. Lodders, H. Palme, and H.-P. Gail, Abundances of the elements in the solar system. *Landolt Börnstein* **4B**, 712 (2009) https://doi.org/10.1007/978-3-540-88055-4_34.
58. J. José, G.M. Halabi, and M.F. El Eid, Synthesis of C-rich dust in CO nova outbursts. *Astron. Astrophys.* **593**, A54 (2016) https://doi.org/10.1051/0004-6361/201628901.
59. J. José, M. Hernanz, S. Amari, K. Lodders, and E. Zinner, The imprint of nova nucleosynthesis in presolar grains. *Astrophys. J.* **612**, 414-428 (2004) https://doi.org/10.1086/422569.
60. E. Anders and N. Grevesse, Abundances of the elements: meteoritic and solar. *Geochim. Cosmochim. Acta* **53**, 197-214 (1989) https://doi.org/10.1016/0016-7037(89)90286-x.
61. P. Hoppe, J. Leitner, M. Pignatari, and S. Amari, New Constraints for Supernova Models from Presolar Silicon Carbide X Grains with Very High $^{26}Al/^{27}Al$ Ratios. *Astrophys. J. Lett.* **943**, L22 (2023) https://doi.org/10.3847/2041-8213/acb157.
62. N. Liu, C.M.O'D. Alexander, B.S. Meyer, et al., Explosive nucleosynthesis in core-collapse Type II supernovae: Insights from new C, N, Si, and Al-Mg isotopic compositions of presolar grains. *Astrophys. J. Lett.* **961**, L22 (2024) https://doi.org/10.3847/2041-8213/ad18c7.
63. N. Liu, N. Dauphas, S. Cristallo, S. Palmerini, and M. Busso, Oxygen and aluminum-magnesium isotopic systematics of presolar nanospinel grains from CI chondrite Orgueil. *Geochim. Cosmochim. Acta* **319**, 296-317 (2022) https://doi.org/10.1016/j.gca.2021.11.022.
64. J. José, A. Coc, and M. Hernanz, Nuclear uncertainties in the NeNa-MgAl cycles and production of $^{22}Na$ and $^{26}Al$ during nova outbursts. *Astrophys. J.* **520**, 347-360 (1999) https://doi.org/10.1086/307445.
65. J. José, A. Coc, and M. Hernanz, Synthesis of intermediate-mass elements in classical novae: From Si to Ca. *Astrophys. J.* **560**, 897-906 (2001) https://doi.org/10.1086/322979.
66. J. Schneider, When will the next T CrB eruption occur? *Research Notes of the American Astronomical Society* **8**, 272 (2024) https://doi.org/10.3847/2515-5172/ad8bba.
67. B.E. Schaefer, The B & V light curves for recurrent nova T CrB from 1842-2022, the unique pre- and post-eruption high-states, the complex period changes, and the upcoming eruption in 2025.5 ± 1.3. *Mon. Not. R. Astron. Soc.* **524**, 3146-3165 (2023) https://doi.org/10.1093/mnras/stad735.
68. L.R. Nittler and C.M.O'D. Alexander, Automated isotopic measurements of micron-sized dust: application to meteoritic presolar silicon carbide. *Geochim. Cosmochim. Acta* **67**, 4961-4980 (2003) https://doi.org/10.1016/s0016-7037(03)00485-x.
69. N. Liu, A. Steele, L.R. Nittler, et al., Coordinated EDX and micro-Raman analysis of presolar silicon carbide: A novel, nondestructive method to identify rare subgroup SiC. *Meteorit. Planet. Sci.* **52**, 2550-2569 (2017) https://doi.org/10.1111/maps.12954.
70. N. Liu, C.M.O'D. Alexander, J. Wang, S. Cristallo, and D. Vescovi. Stellar origins of types Y and Z silicon carbide grains revealed by nickel isotopes. in *56th Lunar and Planetary Science Conference*. The Woodlands, Texas, #2188 (2025).

71. T. Stephan, R. Trappitsch, P. Hoppe, et al., The presolar grain database. I. Silicon carbide. *Astrophys. J. Suppl. Ser.* **270**, 27 (2024) https://doi.org/10.3847/1538-4365/ad1102.
72. T. Stephan and R. Trappitsch, Reliable uncertainties: Error correlation, rotated error bars, and linear regressions in three-isotope plots and beyond. *Int. J. Mass Spectrom.* **491**, 117053 (2023) https://doi.org/10.1016/j.ijms.2023.117053.
73. A.N. Nguyen, L.R. Nittler, C.M.O'D. Alexander, and P. Hoppe, Titanium isotopic compositions of rare presolar SiC grain types from the Murchison meteorite. *Geochim. Cosmochim. Acta* **221**, 162-181 (2018) https://doi.org/10.1016/j.gca.2017.02.026.
74. M. Lugaro, B. Cseh, B. Világos, et al., Origin of large meteoritic SiC stardust grains in metal-rich AGB stars. *Astrophys. J.* **898**, 96 (2020) https://doi.org/10.3847/1538-4357/ab9e74.
75. S. Cristallo, A. Nanni, G. Cescutti, et al., Mass and metallicity distribution of parent AGB stars of presolar SiC. *Astron. Astrophys.* **644**, A8 (2020) https://doi.org/10.1051/0004-6361/202039492.
76. N. Liu, T. Stephan, S. Cristallo, et al., Presolar silicon carbide grains of types Y and Z: their molybdenum isotopic compositions and stellar origins. *Astrophys. J.* **881**, 28 (2019) https://doi.org/10.3847/1538-4357/ab2d27.
77. N. Liu, M. Lugaro, J. Leitner, B.S. Meyer, and M. Schönbächler, Presolar grains as probes of supernova nucleosynthesis. *Space Sci. Rev.* **220**, 88 (2024) https://doi.org/10.1007/s11214-024-01122-w.
78. E. Zinner, S. Amari, R. Guinness, et al., NanoSIMS isotopic analysis of small presolar grains: Search for $Si_3N_4$ grains from AGB stars and Al and Ti isotopic compositions of rare presolar SiC grains. *Geochim. Cosmochim. Acta* **71**, 4786-4813 (2007) https://doi.org/10.1016/j.gca.2007.07.012.
79. F. Gyngard, S. Amari, E. Zinner, and K.K. Marhas, Correlated silicon and titanium isotopic compositions of presolar SiC grains from the Murchison CM2 chondrite. *Geochim. Cosmochim. Acta* **221**, 145-161 (2018) https://doi.org/10.1016/j.gca.2017.09.031.
80. J. José and M. Hernanz, The origin of presolar nova grains. *Meteorit. Planet. Sci.* **42**, 1135-1143 (2007) https://doi.org/10.1111/j.1945-5100.2007.tb00565.x.
81. D. Vescovi, S. Cristallo, M. Busso, and N. Liu, Magnetic-buoyancy-induced mixing in AGB stars: presolar SiC grains. *Astrophys. J. Lett.* **897**, L25 (2020) https://doi.org/10.3847/2041-8213/ab9fa1.
82. N. Liu, T. Stephan, S. Cristallo, et al., Presolar silicon carbide grains of types Y and Z: their strontium and barium isotopic compositions and stellar origins. *The European Physical Journal A* **58**, 216 (2022)
83. J.M. Scalo, On the limiting mass of carbon-oxygen white dwarfs. *Astrophys. J.* **206**, 215-217 (1976) https://doi.org/10.1086/154374.
84. L. Siess, Evolution of massive AGB stars. I. Carbon burning phase. *Astron. Astrophys.* **448**, 717-729 (2006) https://doi.org/10.1051/0004-6361:20053043.
85. C.L. Doherty, P. Gil-Pons, L. Siess, J.C. Lattanzio, and H.H.B. Lau, Super- and massive AGB stars - IV. final fates - initial-to-final mass relation. *Mon. Not. R. Astron. Soc.* **446**, 2599-2612 (2015) https://doi.org/10.1093/mnras/stu2180.
86. L. Ward, C. Iliadis, M. Bose, et al., Impact of thermonuclear reaction rate uncertainties on the identification of presolar grains from classical novae. *Astrophys. J.* **986**, 109 (2025) https://doi.org/10.3847/1538-4357/add47a.
87. A. Evans, Formation and evolution of dust in novae, in *IAU Colloquium 122: Physics of Classical Novae* (Ed. A. Cassatella and R. Viotti (Exec., 253 (1990) https://doi.org/10.1007/3-540-53500-4_133.
88. A. Evans and J.M.C. Rawlings, Dust and molecules in nova environments, in *Classical Novae* (Ed. M.F. Bode and A. Evans (Exec., 308-334 (2008) https://doi.org/10.1017/cbo9780511536168.015.
89. R.D. Gehrz. Infrared and radio observations of classical novae: Physical parameters and abundances in the ejecta. in *Classical Nova Explosions*. AIP, #198-207 (2002).
90. R.D. Gehrz, Infrared studies of classical novae, in *Classical Novae* (Ed. M.F. Bode and A. Evans (Exec., 167-193 (2008) https://doi.org/10.1017/cbo9780511536168.010.
91. R.D. Gehrz, J.W. Truran, R.E. Williams, and S. Starrfield, Nucleosynthesis in classical novae

and its contribution to the interstellar medium. *Publ. Astron. Soc. Pac.* **110**, 3-26 (1998) https://doi.org/10.1086/316107.

92. S.N. Shore and R.D. Gehrz, Photo-ionization induced rapid grain growth in novae. *Astron. Astrophys.* **417**, 695-699 (2004) https://doi.org/10.1051/0004-6361:20034243.
93. A. Choplin, L. Siess, and S. Goriely, Proton ingestion in asymptotic giant branch stars as a possible explanation for J-type stars and AB2 grains. *Astron. Astrophys.* **691**, L7 (2024) https://doi.org/10.1051/0004-6361/202451013.
94. F. Herwig, M. Pignatari, P.R. Woodward, et al., Convective-reactive proton-$^{12}$C combustion in Sakurai's Object (V4334 Sagittarii) and implications for the evolution and yields from the first generations of stars. *Astrophys. J.* **727**, 89 (2011) https://doi.org/10.1088/0004-637x/727/2/89.
95. M. Jadhav, M. Pignatari, F. Herwig, et al., Relics of ancient post-AGB stars in a primitive meteorite. *Astrophys. J. Lett.* **777**, L27 (2013) https://doi.org/10.1088/2041-8205/777/2/l27.

**Table 8.** Multielement Isotopic data for $^{13}$C-rich presolar SiC grains from this study. Reported are 1σ errors. AB grains lacking clear *s*-process signatures shown in Fig. 7 are labeled with *.[7]

| Grain | Type | $^{12}$C/$^{13}$C | $^{14}$N/$^{15}$N | $^{26}$Al/$^{27}$Al (×10$^{-3}$) | $\delta^{29}Si_{28}$ (‰) | $\delta^{30}Si_{28}$ (‰) | $\delta^{46}Ti_{48}$ (‰) | $\delta^{47}Ti_{48}$ (‰) | $\delta^{49}Ti_{48}$ (‰) | $\delta^{50}Ti_{48}$ (‰) | $\delta^{60}Ni_{58}$ (‰) | $\delta^{61}Ni_{58}$ (‰) | $\delta^{62}Ni_{58}$ (‰) | $\delta^{64}Ni_{58}$ (‰) |
|---|---|---|---|---|---|---|---|---|---|---|---|---|---|---|
| M5-A1-0238* | AB2 | 5.55±0.09 | 620±18 | 4.69±0.29 | 103±11 | 62±10 | 49±10 | −21±10 | 93±15 | 43±15 | 46±36 | | 68±60 | 268±109 |
| M5-A1-0415* | AB1 | 2.64±0.06 | 118±3 | 6.47±0.39 | −3±9 | 5±9 | −21±14 | −6±16 | −22±20 | −59±20 | | | | |
| M5-A2-0446 | AB2 | 5.14±0.13 | 1270±212 | | 61±33 | 28±30 | | | | | | | | |
| M5-A2-1678 | AB1 | 3.10±0.07 | 170±7 | 16.15±0.82 | 37±11 | 40±12 | | | | | 58±66 | | 73±144 | −102±255 |
| M5-A3-0164-1* | AB2 | 12.7±0.30 | 1936±360 | 4.12±0.42 | 14±8 | 28±10 | −35±24 | −51±24 | 34±21 | −36±21 | | | | |
| M5-A3-0453 | Nova | 3.56±0.06 | 14±1 | 334.14±3.43 | 9±11 | 152±12 | 34±30 | 28±35 | 82±27 | 49±27 | | | | |
| M5-A3-0600* | AB2 | 9.01±0.15 | 2162±93 | 5.39±0.26 | 37±10 | 24±10 | −30±10 | −42±10 | −7±14 | −24±15 | 140±44 | | 152±79 | 166±142 |
| M5-A3-0816 | AB1 | 2.16±0.03 | 29±1 | 28.26±1.41 | 80±12 | 55±12 | 60±43 | −94±55 | 183±55 | 452±64 | | | | |
| M5-A3-0913 | Nova | 6.63±0.12 | 16±1 | 682.08±3.56 | −96±15 | −14±18 | 46±40 | −19±49 | 193±60 | 104±59 | | | | |
| M5-A4-0663 | AB1 | 2.28±0.04 | 408±66 | | −16±10 | 10±12 | | | | | | | | |
| M5-A4-0678 | AB1 | 2.42±0.04 | 120±3 | 13.96±1.54 | 3±10 | 16±10 | | | | | | | | |
| M5-A4-0685* | AB2 | 4.47±0.07 | 2118±127 | | −25±6 | 12±8 | −31±15 | −29±16 | 15±18 | −101±17 | | | | |
| M5-A4-0756 | AB2 | 11.7±0.26 | 2394±108 | 2.66±0.11 | −1±6 | 23±5 | −52±15 | −52±16 | −33±20 | 78±23 | 212±47 | | 356±91 | 1058±198 |
| M5-A5-0309* | AB1 | 3.03±0.05 | 102±3 | 6.32±0.49 | 23±9 | 27±9 | −25±17 | −43±19 | −22±19 | −109±17 | 81±79 | | −98±159 | 18±333 |
| M5-A5-0697 | AB2 | 8.80±0.18 | 2042±108 | 1.68±0.08 | 0±8 | 28±8 | 12±16 | −22±16 | 65±17 | 43±18 | 124±44 | | 413±93 | 825±185 |
| M5-A5-1064 | AB1 | 3.76±0.02 | 194±6 | 3.92±0.13 | 116±8 | 72±10 | | | | | | | | |
| M5-A5-2546 | AB1 | 2.60±0.04 | 143±6 | 11.79±1.49 | −43±9 | 8±9 | −42±22 | −54±27 | −29±20 | −21±20 | 261±110 | | 745±291 | 1533±682 |
| M5-A5-2774 | AB1 | 1.71±0.03 | 26±1 | 31.05±1.23 | 140±14 | 81±12 | 3±39 | −17±46 | 58±65 | 82±68 | | | | |
| M5-A5-3063 | AB1 | 2.23±0.03 | 209±5 | 6.13±0.39 | −9±10 | 31±6 | −112±12 | −82±12 | −9±16 | 71±17 | 266±99 | | 160±206 | 1023±535 |
| M5-A5-3103* | AB1 | 2.92±0.04 | 95±2 | 7.48±0.19 | 4±10 | 14±7 | −64±14 | −10±15 | 28±24 | −5±24 | | | | |
| M5-A5-3268 | AB2 | 3.27±0.05 | 420±23 | 4.10±0.39 | 8±10 | 24±7 | −52±9 | −69±10 | 36±19 | 158±21 | | | | |
| M5-A5-4012-2 | AB2 | 13.1±0.26 | 7200±836 | 0.92±0.15 | 7±15 | 39±17 | −25±38 | −83±38 | 28±106 | 69±112 | | | | |
| M5-A5-4314 | AB2 | 3.19±0.06 | 873±38 | 12.07±0.66 | 59±11 | 66±11 | 60±83 | −24±83 | 63±23 | 103±24 | 20±51 | | 67±109 | 1053±288 |
| M5-A6-0098 | AB2 | 9.41±0.06 | 5022±879 | | 7±5 | 13±8 | −10±23 | −25±24 | 88±24 | 105±26 | | | | |
| M5-A6-0831 | Nova | 3.90±0.02 | 26±1 | 49.86±1.82 | −6±5 | 65±8 | −48±22 | 2±23 | 20±34 | 81±35 | | | | |
| M5-A6-1067 | AB2 | 13.1±0.09 | 832±114 | | 54±6 | 42±9 | | | | | | | | |
| M5-A7-0123-1 | AB2 | 12.1±0.19 | 1777±83 | 1.94±0.12 | 122±12 | 95±9 | | | | | | | | |
| M5-A7-0757* | AB2 | 8.57±0.14 | 3102±178 | 2.45±0.29 | 2±11 | 37±11 | −12±13 | −16±13 | 24±18 | 45±19 | | | | |
| M5-A7-1189-2* | AB2 | 3.67±0.02 | 3309±237 | 1.61±0.15 | 15±11 | 31±9 | −114±9 | −61±9 | −74±15 | −125±15 | | | | |
| M5-A7-1754 | AB2 | 11.0±0.17 | 4336±191 | 2.04±0.11 | −3±10 | 26±6 | −13±9 | −33±9 | 62±16 | 105±17 | 159±26 | 563±99 | 319±51 | 1099±141 |
| M5-A7-2270-1 | AB1 | 4.00±0.08 | 298±12 | 5.26±0.29 | 173±11 | 153±11 | | | | | | | | |
| M5-A7-3202 | AB1 | 3.87±0.01 | 288±9 | 9.57±1.62 | 170±10 | 138±10 | 119±17 | 15±16 | 183±23 | 267±24 | 11±102 | | 694±311 | 350±538 |
| M6-A1-2909 | AB2 | 11.0±0.10 | 5871±1177 | 1.82±0.28 | 16±6 | 21±7 | 25±38 | −39±38 | 48±27 | 107±29 | | | | |
| M6-A2-0312 | AB1 | 5.40±0.10 | 93±10 | | 30±15 | 66±18 | | | | | | | | |
| M6-A2-0515 | AB2 | 7.50±0.10 | 2871±279 | 4.85±0.22 | 198±7 | 149±7 | 116±19 | 41±9 | 151±17 | 206±18 | −21±45 | | 171±100 | 551±214 |
| M6-A3-0253-1 | AB1 | 2.37±0.01 | 42±1 | 28.42±1.15 | 39±4 | 56±7 | 15±37 | −44±37 | 170±69 | 238±72 | | | | |
| M6-A3-0311 | AB1 | 3.41±0.02 | 59±1 | 21.05±0.68 | −10±5 | 17±8 | | | | | | | | |
| M6-A3-1113* | AB1 | 3.04±0.02 | 96±2 | 20.26±0.84 | −24±3 | −14±6 | −26±14 | −22±15 | 20±24 | 29±24 | 147±53 | | 37±100 | 262±208 |
| M6-A3-1153 | AB2 | 4.10±0.01 | 702±80 | 6.89±0.61 | 41±12 | 43±14 | 18±17 | −34±17 | 23±17 | 47±18 | 130±85 | | 582±225 | −352±276 |
| M6-A3-1240-2 | AB2 | 4.87±0.03 | 5201±1346 | 9.76±2.10 | −7±5 | 3±8 | | | | | | | | |

[7]For delta notation, the isotope ratios for normalization are 0.506 ($^{29}$Si/$^{28}$Si), 0.0334 ($^{30}$Si/$^{28}$Si), 0.1085 ($^{46}$Ti/$^{48}$Ti), 0.1008 ($^{47}$Ti/$^{48}$Ti), 0.0745 ($^{49}$Ti/$^{48}$Ti), 0.0734 ($^{50}$Ti/$^{48}$Ti), 0.3857 ($^{60}$Ni/$^{58}$Ni), 0.0167 ($^{61}$Ni/$^{58}$Ni), 0.0535 ($^{62}$Ni/$^{58}$Ni), and 0.0256 ($^{64}$Ni/$^{58}$Ni).

| M6-A3-1313* | AB2 | 11.0±0.10 | 2025±371 | 4.41±0.54 | −37±10 | −50±12 | −34±17 | −77±18 | −60±16 | −133±16 | 349±74 |  | 196±141 | 555±308 |
|---|---|---|---|---|---|---|---|---|---|---|---|---|---|---|
| M6-A3-1679 | AB2 | 6.90±0.10 | 2456±492 | 1.93±0.34 | −1±7 | 41±8 | 61±17 | −42±16 | 107±19 | 311±22 | 199±57 | 224±362 | 897±260 | 2918±471 |
| M6-A3-1877 | Nova | 10.8±0.05 | 24±1 | 315.38±21.01 | 57±10 | 129±12 | −21±24 | −54±28 | 120±28 | 62±27 |  |  |  |  |
| M6-A3-2005* | AB1 | 4.60±0.01 | 381±12 | 9.67±0.72 | 63±7 | 44±8 | −17±11 | −21±12 | 44±11 | −14±11 | 48±36 | 416±170 | 128±195 | 582±195 |
| M6-A3-2252 | AB2 | 11.4±0.10 | 2939±930 |  | 26±8 | 35±9 |  |  |  |  |  |  |  |  |
| M6-A4-0523 | AB2 | 11.1±0.05 | 2853±285 | 1.17±0.32 | 0±7 | 16±5 | −19±15 | −19±15 | 59±22 | 32±22 | 96±136 | 759±344 | 454±177 |  |
| M6-A4-0789 | AB2 | 9.60±0.10 | 1215±184 | 3.58±0.59 | 56±9 | 47±11 | 19±15 | −29±16 | 51±13 | 29±13 | 48±84 |  | 217±205 | 1579±589 |
| M6-A4-1471 | AB1 | 4.79±0.02 | 140±10 | 2.81±0.40 | −11±9 | 6±10 |  |  |  |  |  |  |  |  |
| M6-A4-1961-1* | AB1 | 2.48±0.02 | 43±1 | 32.10±1.03 | 45±5 | 56±7 | −32±8 | −38±8 | −11±8 | 8±9 | 19±63 |  | 162±147 | 289±298 |
| M6-A4-2768 | AB2 | 6.21±0.03 | 1044±90 | 2.83±0.35 | 36±9 | 34±11 |  |  |  |  |  |  |  |  |
| M6-A4-2834 | AB2 | 3.72±0.02 | 418±23 |  | −25±6 | 12±8 | −39±34 | −1±37 | 41±34 | 85±36 |  |  |  |  |
| M6-A4-3243-a | AB1 | 4.00±0.01 | 349±38 |  | 173±7 | 135±9 | 137±21 | 80±21 | 135±25 | 172±27 |  |  |  |  |
| M6-A4-3471 | AB1 | 3.20±0.01 | 91±4 | 28.76±1.51 | 133±7 | 120±8 | 147±28 | 66±28 | 75±25 | 277±28 | 39±82 |  | 414±218 | 1796±600 |
| M6-A4-3905 | AB1 | 3.07±0.03 | 198±15 |  | 24±7 | 24±8 | 53±19 | −17±19 | 102±27 | 135±28 |  |  |  |  |
| M6-A5-0149 | AB1 | 3.10±0.01 | 112±4 |  | −12±6 | −2±7 |  |  |  |  | 101±58 |  | 172±125 | 590±277 |
| M6-A5-0163-2 | AB2 | 11.8±0.11 | 13563±1128 | 0.89±0.08 | 34±5 | 47±6 | 13±14 | −20±14 | 50±43 | 113±46 | 95±46 |  | 132±90 | 1573±253 |
| M6-A5-0582 | AB2 | 5.88±0.02 | 467±37 |  | 47±9 | 30±10 |  |  |  |  |  |  |  |  |
| M6-A5-0982 | AB1 | 2.57±0.03 | 112±4 |  | −41±12 | −7±15 |  |  |  |  |  |  |  |  |
| M6-A5-1000-1 | AB2 | 5.76±0.03 | 3795±328 | 0.50±0.05 | 8±8 | 31±7 |  |  |  |  | 112±46 |  | 187±89 | 2007±260 |
| M6-A5-1210 | AB1 | 3.39±0.02 | 59±1 | 15.03±0.50 | −7±5 | 19±7 | −20±19 | −24±19 | 67±22 | 264±25 |  |  |  |  |
| M6-A6-0336 | AB1 | 4.54±0.03 | 337±60 |  | 5±7 | 40±10 |  |  |  |  |  |  |  |  |
| M6-A6-0413 | AB1 | 3.36±0.02 | 57±1 | 14.66±0.50 | −29±10 | −15±13 |  |  |  |  |  |  |  |  |
| M6-A6-0645-1 | AB1 | 2.59±0.02 | 75±3 | 18.67±1.31 | −7±8 | −6±11 |  |  |  |  | 329±92 |  | 109±179 | 839±451 |
| M6-A6-0703* | AB2 | 4.01±0.03 | 3891±752 | 2.55±0.25 | 62±7 | 43±10 | −12±22 | −1±23 | 100±24 | 44±24 |  |  |  |  |
| M6-A6-0794* | AB2 | 10.8±0.07 | 750±121 | 1.93±0.14 | 41±5 | 32±8 | 4±22 | −27±23 | 18±24 | −25±25 |  |  |  |  |
| M6-A6-0913 | AB1 | 2.04±0.01 | 71±2 |  | 0±6 | 9±9 |  |  |  |  |  |  |  |  |
| M6-A6-0987 | AB2 | 8.13±0.05 | 526±95 |  | 113±6 | 102±9 |  |  |  |  |  |  |  |  |
| M6-A6-1376 | AB2 | 6.55±0.04 | 1995±229 | 0.55±0.09 | −3±4 | 39±7 | −34±18 | −43±19 | 24±17 | 580±23 | 435±89 |  | 63±282 | 1354±504 |
| M6-A6-1679 | AB2 | 6.81±0.03 | 2766±546 | 1.47±0.15 | 5±7 | 36±8 |  |  |  |  |  |  |  |  |
| M6-A6-1719 | AB1 | 4.45±0.03 | 232±8 | 2.68±0.27 | 111±4 | 84±7 | 40±58 | 67±61 | 99±36 | 117±37 |  |  |  |  |
| M6-A6-2021 | AB1 | 4.53±0.03 | 393±28 | 4.93±0.27 | 73±4 | 42±7 | 4±14 | −7±14 | 49±18 | −47±18 | 77±71 |  | −11±149 | 1465±464 |
| M6-A6-2082-2 | AB2 | 7.94±0.05 | 433±69 |  | −24±6 | 22±9 |  |  |  |  |  |  |  |  |
| M6-A6-2346 | AB2 | 6.18±0.04 | 4019±493 | 1.53±0.10 | 30±5 | 56±8 | 10±16 | −27±16 | 94±16 | 169±18 |  |  |  |  |
| M6-A6-2899* | AB1 | 3.14±0.02 | 100±2 | 11.42±0.60 | −24±6 | −13±9 | −15±34 | 62±37 | 8±46 | 34±48 |  |  |  |  |
| M6-A6-3018 | AB2 | 8.38±0.05 | 7021±772 | 1.39±0.09 | 10±5 | 35±8 | −28±15 | −31±16 | 79±20 | 98±21 | 239±71 |  | 387±159 | 1441±404 |
| M6-A6-3241 | AB2 | 10.8±0.07 | 2234±708 |  | 103±12 | 74±14 |  |  |  |  |  |  |  |  |
| M6-A6-3271-1 | AB2 | 9.51±0.05 | 596±42 |  | 30±10 | 35±12 |  |  | 40±26 | 147±28 |  |  |  |  |
| M6-A6-3362* | AB2 | 6.03±0.04 | 8569±776 | 1.54±0.09 | −14±4 | 1±6 | −85±36 | −29±39 | 28±41 | 14±42 |  |  |  |  |
| M6-A7-2579* | AB1 | 3.05±0.03 | 44±1 | 36.66±2.97 | 13±6 | 6±7 |  |  | 92±79 | 28±77 |  |  |  |  |
| M6-A7-2645 | AB1 | 3.45±0.03 | 396±14 | 1.78±0.17 | −44±9 | −50±10 | −13±22 | −62±23 | 16±22 | 164±25 |  |  |  |  |
| M6-A7-3129 | AB2 | 14.0±0.10 | 3331±683 |  | 25±7 | 32±9 | 28±28 | 16±29 | 138±22 | 170±24 |  |  |  |  |
| M6-A7-3214 | AB2 | 12.9±0.05 | 4283±788 | 0.82±0.11 | −36±6 | −19±6 |  |  |  |  |  |  |  |  |
| M6-A7-3406 | AB2 | 7.93±0.05 | 4779±1237 | 1.04±0.17 | 30±6 | 48±9 |  |  |  |  |  |  |  |  |